\documentclass[%
 reprint,
superscriptaddress,
nofootinbib,
 amsmath,amssymb,
 prl,
floatfix,
]{revtex4-2}
\usepackage{amsmath}
\usepackage{amssymb}
\usepackage{mathtools}
\usepackage[tight]{subfigure}
\usepackage{enumitem}
\usepackage{soul}
\usepackage{cancel}
\usepackage{multirow}
\usepackage{tikz}
\usepackage{pifont}
\usepackage{xspace}
\usepackage{comment}
\usepackage{physics}
\usepackage{bbold}
\usepackage{graphicx}
\usepackage{dcolumn}
\usepackage{bm}
\usepackage[breaklinks]{hyperref}
\hypersetup{colorlinks=true, linkcolor=blue, citecolor=blue, filecolor=blue, urlcolor=blue}
\usepackage{ulem}

\usepackage[capitalise]{cleveref}

\begin{document}


\title{Entanglement Steering in Adaptive Circuits with Feedback}
\author{Vikram Ravindranath}
\affiliation{Department of Physics, Boston College, Chestnut Hill, MA 02467, USA}

\author{Yiqiu Han}
\affiliation{Department of Physics, Boston College, Chestnut Hill, MA 02467, USA}

\author{Zhi-Cheng Yang}
\email{zcyang19@pku.edu.cn}
\affiliation{School of Physics, Peking University, Beijing 100871, China}
\affiliation{Center for High Energy Physics, Peking University, Beijing 100871, China}

\author{Xiao Chen}
\affiliation{Department of Physics, Boston College, Chestnut Hill, MA 02467, USA}

\date{\today}

\begin{abstract}

The intensely studied measurement-induced entanglement phase transition has become a hallmark of non-unitary quantum many-body dynamics. Usually, such a transition only  appears at the level of each individual quantum trajectory, and is absent for the density matrix averaged over measurement outcomes. In this work, we introduce a class of adaptive random circuit models with feedback that exhibit transitions in both settings. After each measurement, a unitary operation is either applied or not depending on the measurement outcome, which steers the averaged density matrix towards a unique state above a certain measurement threshold. Interestingly, the  transition for the density matrix and the entanglement transition in the individual quantum trajectory in general happen at \textit{different} critical measurement rates. We demonstrate that the former transition belongs to the parity-conserving universality class by explicitly mapping to a classical branching-annihilating random walk process.

\end{abstract}

\maketitle


\textit{Introduction.-} Monitoring a quantum system can yield fascinating physics. The interplay between unitary dynamics and measurement creates an intriguing non-equilibrium phenomenon referred to as measurement-induced entanglement phase transitions (MIPTs)~\cite{PhysRevB.98.205136, PhysRevB.100.134306, PhysRevX.9.031009, PhysRevB.99.224307,Gullans_2020,Choi_2020}. In its most commonly explored setting, an initial unentangled state is evolved by a random quantum circuit subject to measurements at a rate $p$. Above (below) a critical rate $p_c$, the steady-state exhibits area-law (volume-law) entanglement scaling. To observe this transition, it is necessary to keep track of each individual quantum trajectory as the intrinsic randomness of measurement outcomes leads to the ensemble-averaged post-measurement state being a maximally mixed density matrix ($\rho \propto \mathbb{1}$) in the allowed Hilbert space. As a result, the MIPT remains invisible to the ensemble-averaged density matrix, presenting significant challenges for experimental observation, with only a few exceptions~\cite{PhysRevLett.126.060501, noel2022measurement,Koh_2022}.

Despite these challenges, mid-circuit repeated measurements have become a valuable tool to create novel quantum phases dynamically. Recently, a new class of non-unitary dynamics has been proposed, where the outcome of a measurement can impact the dynamics themselves, leading to a non-trivial density matrix and stabilizing various quantum ordered phases through a feedback mechanism~\cite{McGinley_2022,Friedman_2022,Iadecola_2022,buchhold2022revealing,iqbal2023topological,foss2023experimental}.

\begin{figure}[!t]
\includegraphics[width=.45\textwidth]{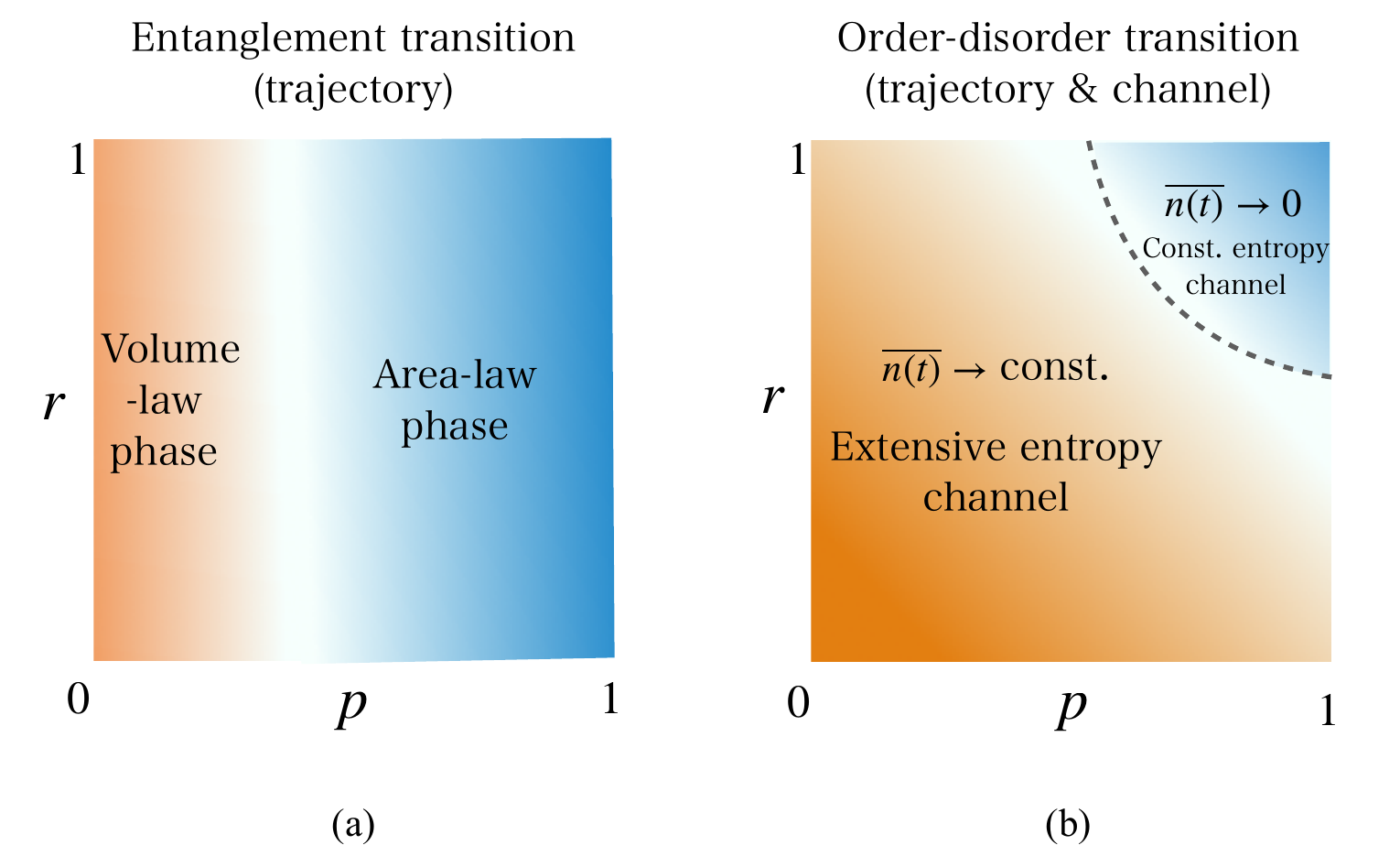}
\caption{Schematic phase diagram illustrating (a) the entanglement phase transition for the individual quantum trajectories and (b) domain wall density transition as a function of the measurement rate $p$ and feedback rate $r$. The transition in (b) also signals an entropy transition \textit{at the level of quantum channel} (ensemble-averaged density matrix).
The critical line in (b) belongs to the PC universality class and satisfies $p\times r={\rm const.}(\approx 0.55)$. The critical line in (a) is numerically upper bounded by $p_c^{EE}\leq 0.45$.} 
\label{fig:phase_diagram} 
\end{figure}

In this Letter, we introduce a class of adaptive random circuits with feedback that exhibits phase transitions for both the quantum trajectory and the ensemble-averaged density matrix (i.e., quantum channel), as depicted schematically in Fig.~\ref{fig:phase_diagram}. In addition to varying the measurement rate $p$, the post-measurement state is also locally corrected conditioned on the measurement outcome with a feedback rate $0\leq r \leq 1$. The feedback is designed to ``steer" the system towards particular final states. When $p\times r$ is large enough, the steady state is a mixture of two ferromagnetically ordered states instead of a maximally mixed state involving exponentially many configurations. 
Thus, there is an order-disorder phase transition in the quantum dynamics, which can be observed at the level of both quantum trajectory and quantum channel~\cite{Iadecola_2022,buchhold2022revealing}. A similar order-disorder transition has also been discussed recently in the context of dissipative phase transitions \cite{sierant2022dissipative, Lang2015dissipative}. However, we emphasize that the transitions discovered therein only exist in $d$-dimension with $d\geq 2$ or $1d$ systems with long-range interactions. Instead, by explicitly mapping the motion of domain walls to a classical branching-annihilating random walk (BAW) process, we show that the order-disorder phase transition in our adaptive circuit model belongs to the parity-conserving (PC) universality class which exists in $1d$~\cite{henkel2008non, zhong1995universality, PhysRevE.63.065103, PhysRevB.105.064306}. Furthermore, the familiar MIPT is observed at the level of the quantum trajectory. Interestingly, we find that these two transitions typically occur at \textit{different} critical measurement rates. Moreover, we show that the order-disorder phase transition can be probed experimentally from the measurement outcomes along the circuit evolution, and does not require post-selection.
\begin{figure}[!t]
\includegraphics[width=.45\textwidth]{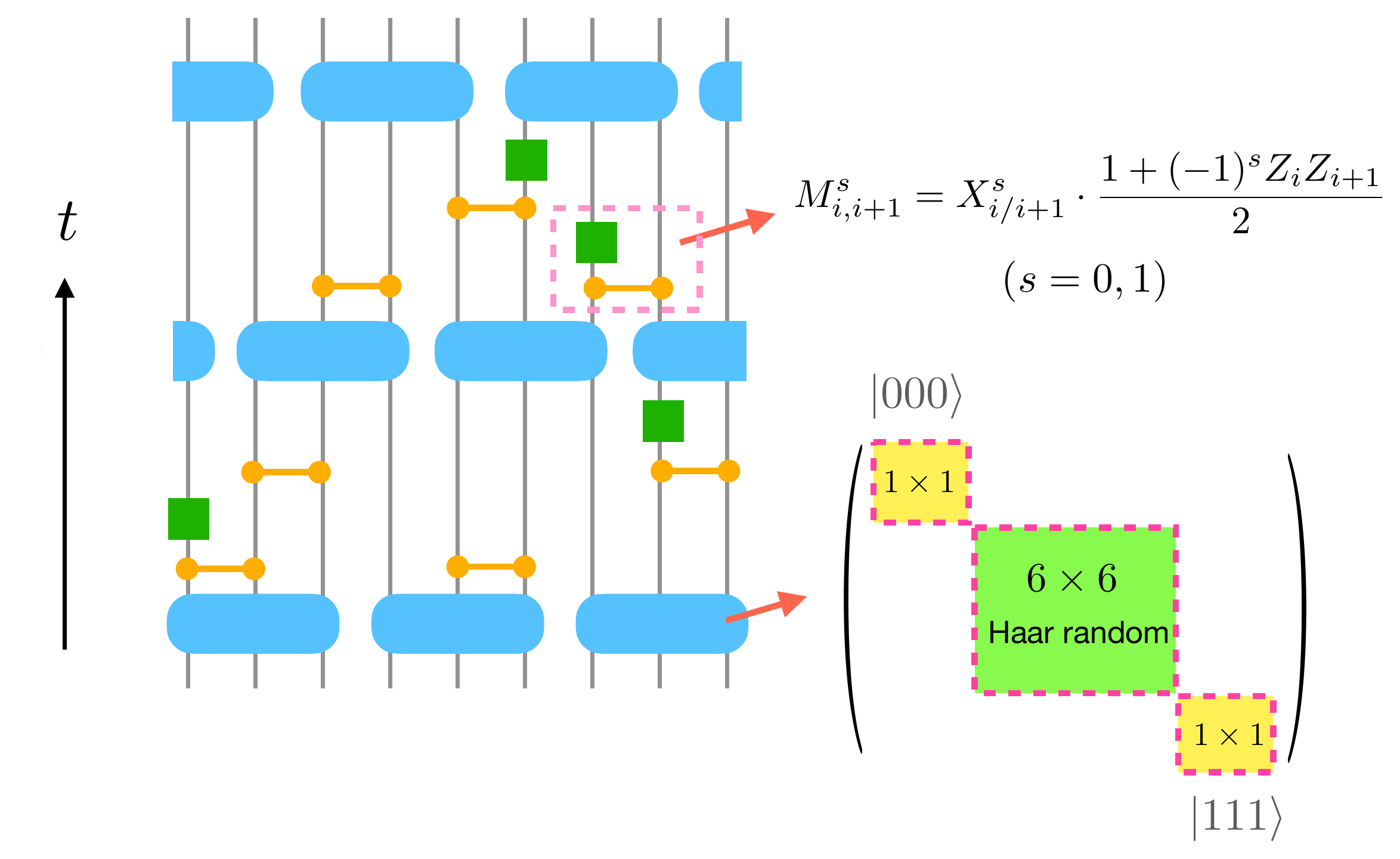}
\caption{Setup of the adaptive circuit with feedback (one time unit). The circuit consists of three-qubit unitary gates (denoted by blue boxes) and two-qubit measurements (orange lines with dotted endpoints). The unitary gate leaves $|000\rangle$ and $|111\rangle$ unchanged, and acts as a Haar random unitary within the complementary subspace. If the measurement outcome $Z_iZ_{i+1}=-1$, a Pauli-$X$ operator (green box) is applied on one site to correct the post-measurement state into $|11\rangle$ or $|00\rangle$. Each unit time step contains three layers of unitary gates related by translation by one site, and two sets of measurements. Each set consists of an even and an odd layer, forming a brickwork structure. }
\label{fig:model} 
\end{figure}

\textit{Model.-} Our circuit model consists of three-qubit unitary gates and two-qubit measurements, as illustrated in Fig.~\ref{fig:model}. We take each three-qubit unitary gate to have a block structure: it leaves the basis states $|000\rangle$ and $|111\rangle$ unchanged up to a U(1) phase (forming blocks of size one), and acts as a Haar random unitary within the six-dimensional subspace spanned by $\{|001\rangle, |010\rangle, |100\rangle, |011\rangle, |101\rangle, |110\rangle \}$. We measure the product of Pauli-$Z$ operators across a bond between two consecutive sites $Z_i Z_{i+1}$.  Crucially, our construction includes \textit{feedback}: if an \lq\lq{}undesired\rq\rq measurement outcome $Z_iZ_{i+1}=-1$ is obtained (meaning that the post-measurement state is $|01\rangle$ or $|10\rangle$), we apply a Pauli-$X$ operator on either site with probability $r$ to correct the state into $|00\rangle$ or $|11\rangle$; if the measurement outcome is +1, then no operator is applied. A set of measurements and corrections is implemented in two layers -- first over odd and then even bonds. In each layer, measurements are made at a rate $p$ per bond. Each unit time step contains three layers of unitary gates related by a translation by one site, interspersed with a set of measurements after each of the first and second, but not third, layers. Time evolution for each unnormalized quantum trajectory is thus given by:
\begin{eqnarray}
    |\psi(\{\bm s\},T)\rangle &=&\prod_{t=1}^T \left[ U_3(t) M_2(t) U_2(t) M_1(t) U_1(t)\right] \ |\psi_0\rangle \nonumber \\
    &\equiv& C(\{\bm s\})\ |\psi_0\rangle,
\label{eq:trajectory}
\end{eqnarray}
where  $U_j$ denotes the $j^\text{th}$ layer of unitary gates at each time step, $M_{1/2}$ denotes a set of measurements, and $\{\bm s\}$ records the full set of measurement outcomes along this particular trajectory.

Akin to previous studies on measurement-induced phase transitions in hybrid random circuits, we expect that there is an MIPT in our setup for individual quantum trajectories [Fig.~\ref{fig:phase_diagram}(a)]. Upon increasing the measurement rate $p$, there is a transition in the time-evolved states~(\ref{eq:trajectory}) from a volume-law entangled phase to a weakly entangled area-law phase, where the spins almost point along the $\hat{z}$ direction. Without feedback $(r=0)$, each spin could randomly point along $\pm \hat{z}$ directions in each trajectory, depending on the measurement outcomes. Upon introducing feedback, the final states in all trajectories will approach a ferromagnetically ordered state $\alpha |00\ldots 0\rangle + \beta |11\ldots1\rangle$, provided that the measurement and feedback rates are large enough. This indicates that besides the MIPT, the steady state also exhibits an order-disorder transition in a physical observable -- the domain wall density -- due to feedback [Fig.~\ref{fig:phase_diagram}(b)]\footnote{In our model, the steady state persists until a timescale $t\sim {\rm poly}(L)$ with a power $\geq 2$. For $t\sim {\rm exp}(L)$, the states will eventually approach a ferromagnetically ordered state, and both the entanglement and order-disorder phase transitions no longer exist.}. Since the expectation value of a physical observable is linear in the density matrix, the average domain wall density evaluated for the trajectories and the quantum channel (trajectory-averaged density matrix)
\begin{equation}
    \rho = \sum_{\{\bm s\}} C(\{\bm s \})\ |\psi_0\rangle \langle \psi_0| \ C(\{\bm s \})^\dagger
\label{eq:density_matrix}
\end{equation}
are equivalent. Hence, this order-disorder transition can be observed both in the individual trajectories and the quantum channel.


\begin{figure}[!t]
\subfigure[]{
\includegraphics[width=.45\textwidth]{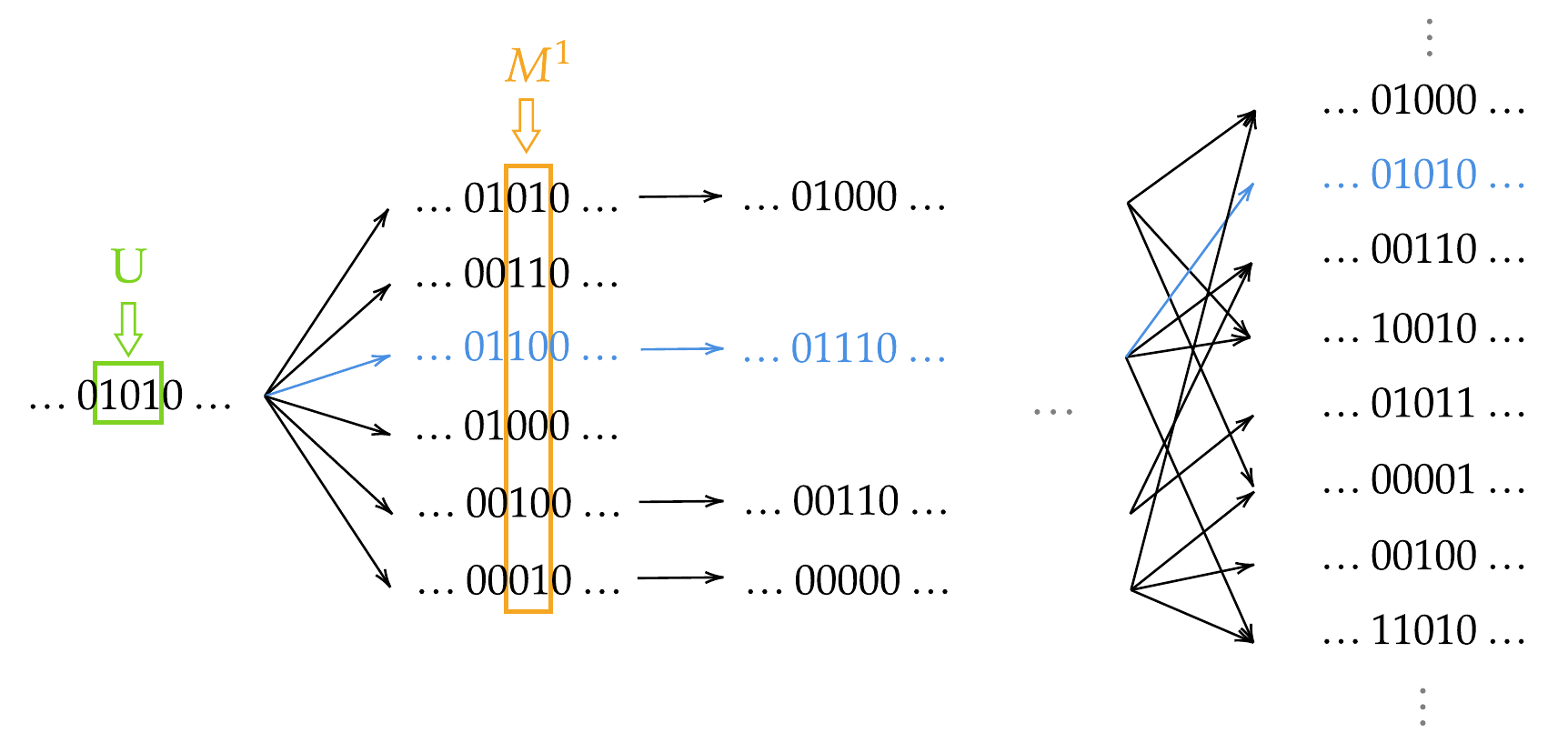}
}

\subfigure[]{
\includegraphics[width=.45\textwidth]{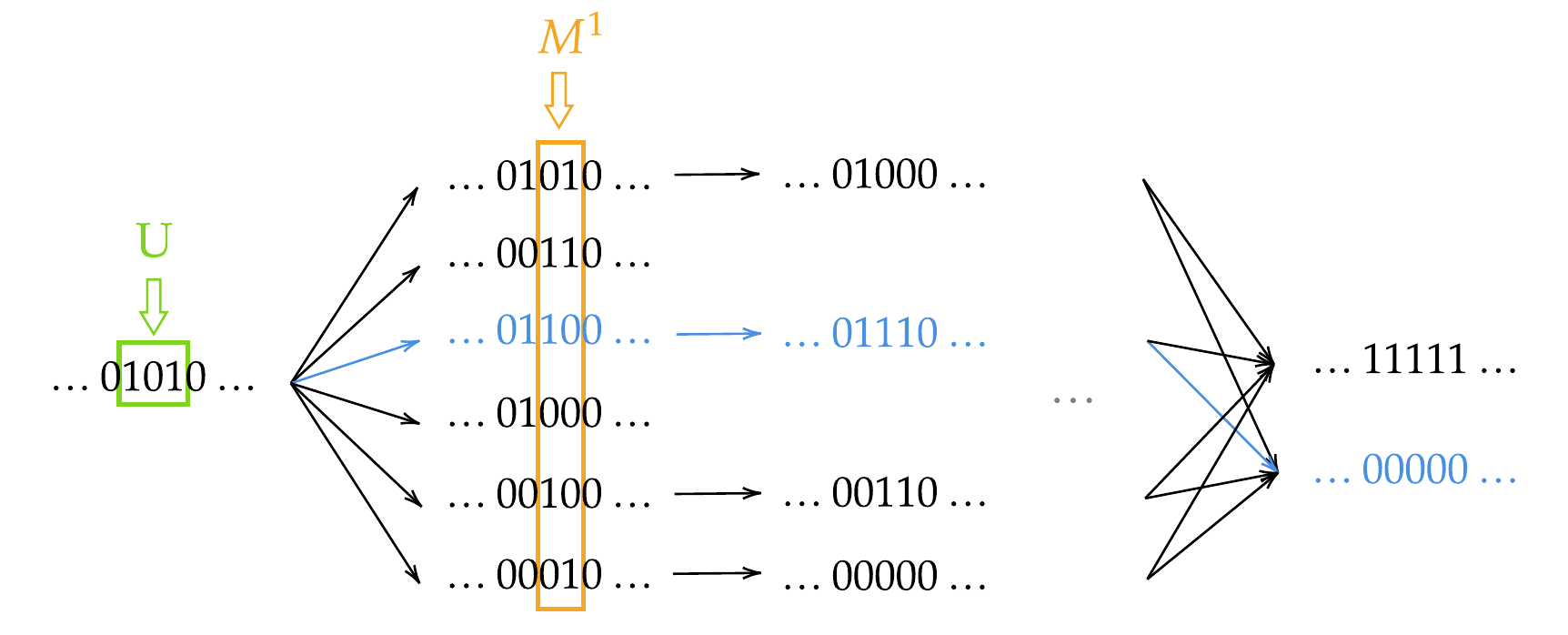}
}

\subfigure[]{
\includegraphics[width=.45\textwidth]{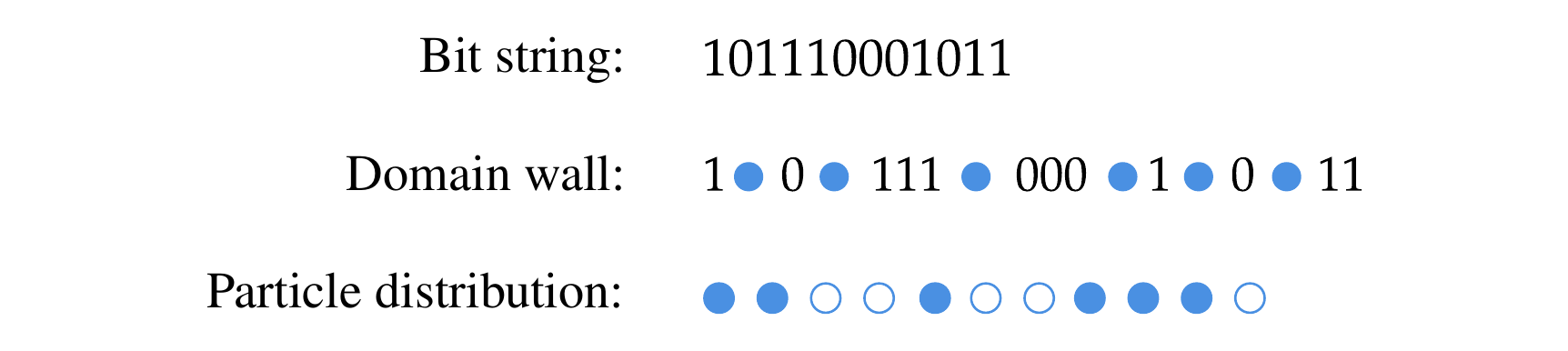}
}
\caption{(a)\&(b): Illustration of the bitstring dynamics in the quantum trajectory undergoing hybrid circuit evolution. The blue arrows indicate two representative paths (``world histories") of the bitstring dynamics. (a) When the measurement rate is small, the steady state involves exponentially many bistring configurations. (b) When the measurement rate is high, the steady state is spanned by two ordered bitstrings. (c) Mapping from a bitstring configuration to particle distribution, where a particle and an empty site represent the presence and absence of a domain wall, respectively.} 
\label{fig:path} 
\end{figure}

\textit{Phase transition in the domain wall density.-} We demonstrate that there is indeed an order-disorder phase transition captured by the domain wall density by mapping the dynamics of domain walls under the hybrid circuit evolution to a classical stochastic process. We write the time-evolved wavefunction as a sum of world histories: $|\psi(t)\rangle = \sum\limits_{ \{m(\tau)\}} A(\{m(\tau) \}) \ |m\rangle$, where $\{|m\rangle \}$ are bitstrings in the computational basis, and $A(\{m(\tau)\})$ is the amplitude of a particular world history $\{m(\tau)\}_{0\leq \tau \leq t}$. We consider each world history by constructing paths connecting bitstring configurations in the initial state to state at time $t$, as illustrated in Fig.~\ref{fig:path}(a)\&(b).

Each individual path can be mapped to a classical stochastic process, which has a non-equilibrium phase transition. We denote the presence of each domain wall as a particle $\bullet$, and the absence of which as an empty site $\circ$. The bitstring configuration is thus translated into one of particle occupations, as shown in Fig.~\ref{fig:path}(c). Under the action of unitary gates, the particles undergo two types of processes: $\bullet \circ \circ \leftrightarrow \bullet \bullet \bullet$ (branching), and diffusion. Under measurement, the particles can either diffuse: $\bullet \circ \leftrightarrow \circ \bullet$, or annihilate in pairs with probability $q\equiv pr$: $\bullet \bullet \rightarrow \circ \circ$. Combining these processes together, the particles perform BAW with an even number of offspring:
\begin{equation}
    W \leftrightarrow 3W, \quad W + W \xrightarrow[]{q} \emptyset.
\end{equation}
Since the parity of the total particle number is conserved, the classical dynamics described above belongs to the PC universality class, which has a continuous dynamical phase transition when the rate of particle annihilation exceeds a certain threshold~\cite{henkel2008non, zhong1995universality, PhysRevE.63.065103, PhysRevB.105.064306} [see Supplemental Material (SM)~\cite{SM} for more details]. The two phases can be distinguished in terms of the average particle (domain wall) density in the steady state $n(t)=N(t)/L$.

For initial conditions with an extensive number of particles $N\propto L$, $n(t)$ saturates to a finite constant when $q < q_c$ after a finite amount of time.
When $q>q_c$, $n(t) \sim t^{-1/2}$, and decays to zero at $t\sim L^2$ if the initial state has an even number of particles. At $q_c$, $n(t) \sim t^{-\theta}$ with a universal exponent $\theta=0.286$ characteristic of PC universality class. 
For the initial condition with two adjacent particles, when $q<q_c$, the particle density grows linearly in time and saturates to a finite constant. At $q=q_c$, the particle density remains constant. When $q>q_c$, the particle density decays as $t^{-1/2}$. Therefore, our model exhibits an order-disorder phase transition at a critical effective measurement rate $q_c$. Below, we shall explicitly demonstrate both transitions depicted in Fig.~\ref{fig:phase_diagram} via numerical simulations of the hybrid circuit dynamics.

\begin{figure}[!t]
\includegraphics[width=0.475\textwidth]{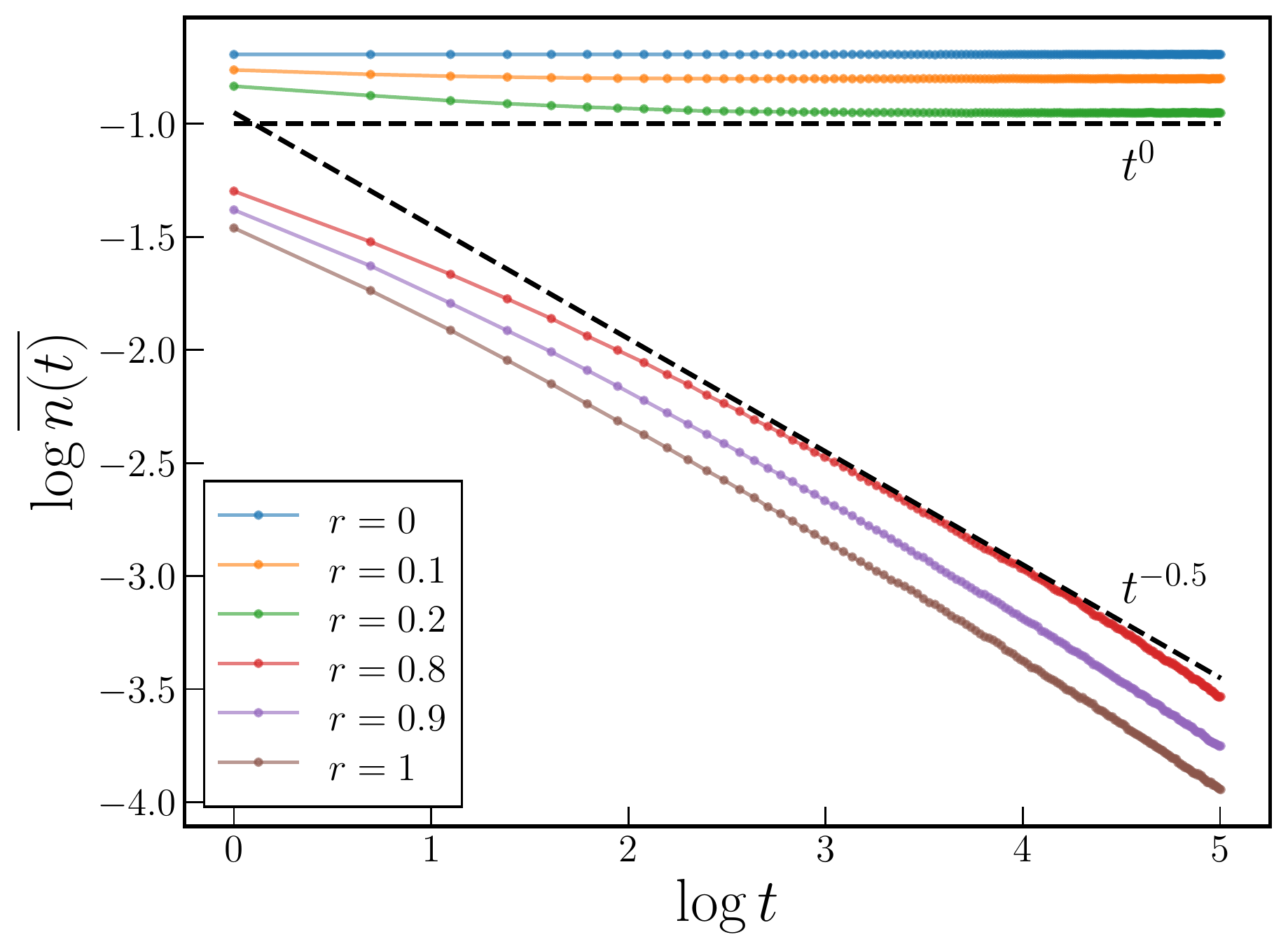}
\caption{The trajectory-averaged domain wall density $\overline{n(t)}$ as a function of time, for different values of the feedback rate $r$ while fixing $p=1$, starting from random initial states. The result confirms the existence of two phases. Numerical simulations are performed for $L=300$, and averaged over $10^4$ realizations of circuits and initial states.}

\label{fig:2LM_RandIni} 
\end{figure}

\textit{Numerical results.-} In our simulations, initial states, which are product states in the computational basis, are evolved according to the setup in \cref{fig:model} with open boundary conditions.
We record, at each time step, the second R\'enyi entropy $S^{(2)}_A(t) = -\log\qty(\text{tr} {\rho^2_A})$, 
where $\rho_A$ denotes the reduced density matrix of subsystem $A$ composed of sites $1, 2\dots |A|; \ 1\leq |A|\leq \frac{L}{2}$, and the domain wall density \begin{align}
    n(t) \equiv \frac{1}{L-1} \sum\limits_{i=1}^{L-1} \expval{\frac{1 - Z_i Z_{i+1}}{2}}_t.
\end{align}
We then compute the trajectory-averaged second R\'enyi entropy $\overline{S^{(2)}_A(t)}$ and domain wall density $\overline{n(t)}$. It is worth noting that in experimental platforms, the determination of $\overline{n}$ does not require an extra step; the measurement outcomes for any $p>0$ provide an accurate sampling of $n(t)$. Our simulations are performed using the ITensor Julia package~\cite{ITensor1,ITensor2} based on a matrix product state representation of the time-evolved wavefunctions.

\begin{figure}[!t]
\includegraphics[width=0.475\textwidth]{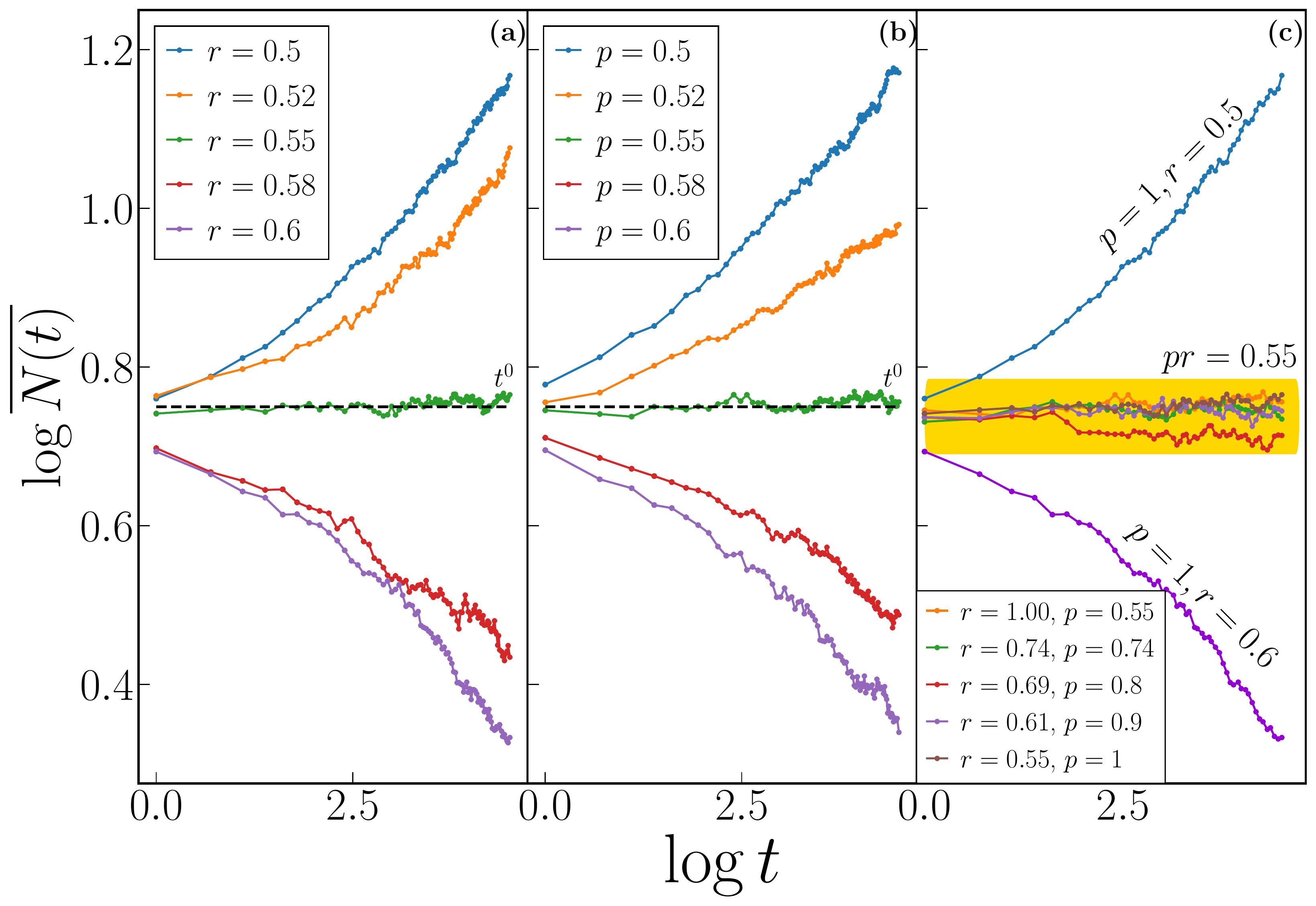}
\caption{The averaged number of domain walls $\overline{N(t)}$ starting from an initial state with two domain walls. (a) For $p=1$, the exponent $\theta$ as in $\overline{n(t)}\sim t^\theta$ changes sign at $r_c\approx0.55$. (b) For $r=1$, $p^n_c\approx0.55$. (c) Demonstration that the order-disorder transition in $\overline{n(t)}$ is governed by a single parameter $q\equiv pr$.}
\label{fig:2LM_qc} 
\end{figure}

We first set $p=1$, and study $\overline{n(t)}$ as $r$ is varied (along the right boundary of Fig.~\ref{fig:phase_diagram}). Since the system is expected to be in the area-law phase, this serves as a benchmark to observe the order-disorder phase transition as revealed by the domain wall density $\overline{n(t)}$. Starting from a random state with $n(t=0)\approx \frac{1}{2}$, we find clear evidence that the steady state is area law entangled and is independent of $r$ (data not shown). Further study of $\overline{n(t)}$ indicates that when $r\lesssim 0.2$, $\overline{n(t)}$ saturates to a finite constant at long times; on the other hand, when $r\gtrsim 0.8$, $\overline{n(t)}\sim t^{-1/2}$ as shown in Fig.~\ref{fig:2LM_RandIni}. This indicates the existence of two phases within the area-law entangled phase, distinguished by $\overline{n(t)}$. However, owing to finite-size effects, $\overline{n(t)}$ appears to decay with a continuously varying exponent (i.e. $n(t)\sim t^{-\theta_{n_0=0.5}(r)}$) for $0.2\lesssim r \lesssim 0.8$, which makes identifying the critical $r_c$ difficult. Instead, we consider an initial state with two neighboring domain walls centered in the lattice. Again, $n(t)\sim t^{\theta_{N_0=2}(r)}$ with a continuously varying exponent in the intermediate regime, but the critical point $r_c$ is determined by the rate $r$ at which $\overline{n(t)} \sim {\rm const.}$ (or equivalently, when $\theta_{N_0=2}(r_c)\approx 0$ and changes sign). Using this criterion, we find that $r_c \approx 0.55$ in \cref{fig:2LM_qc}(a).

Next, we study the dynamics upon varying $p$, while fixing $r=1$ (along the top boundary of Fig.~\ref{fig:phase_diagram}). In this case, we expect to observe an entanglement phase transition \textit{and} a transition in $\overline{n(t)}$. In Fig.~\ref{fig:2LM_qc}(b), we find an order-disorder transition as revealed by $\overline{n(t)}$ at $p_c^n\approx 0.55$. In fact, we find, by considering various values of $p$ and $r$, that the order-disorder transition in our system is controlled by a single parameter $q\equiv pr$, and happens at $q_c\approx 0.55$ (see Fig.~\ref{fig:2LM_qc}(c)), confirming our mapping to a classical BAW process in Fig.~\ref{fig:path}. An interesting question that naturally arises is whether the location of the order-disorder transition in $\overline{n(t)}$ coincides with that of the MIPT.

\begin{figure}[!t]
\includegraphics[width=0.475\textwidth]{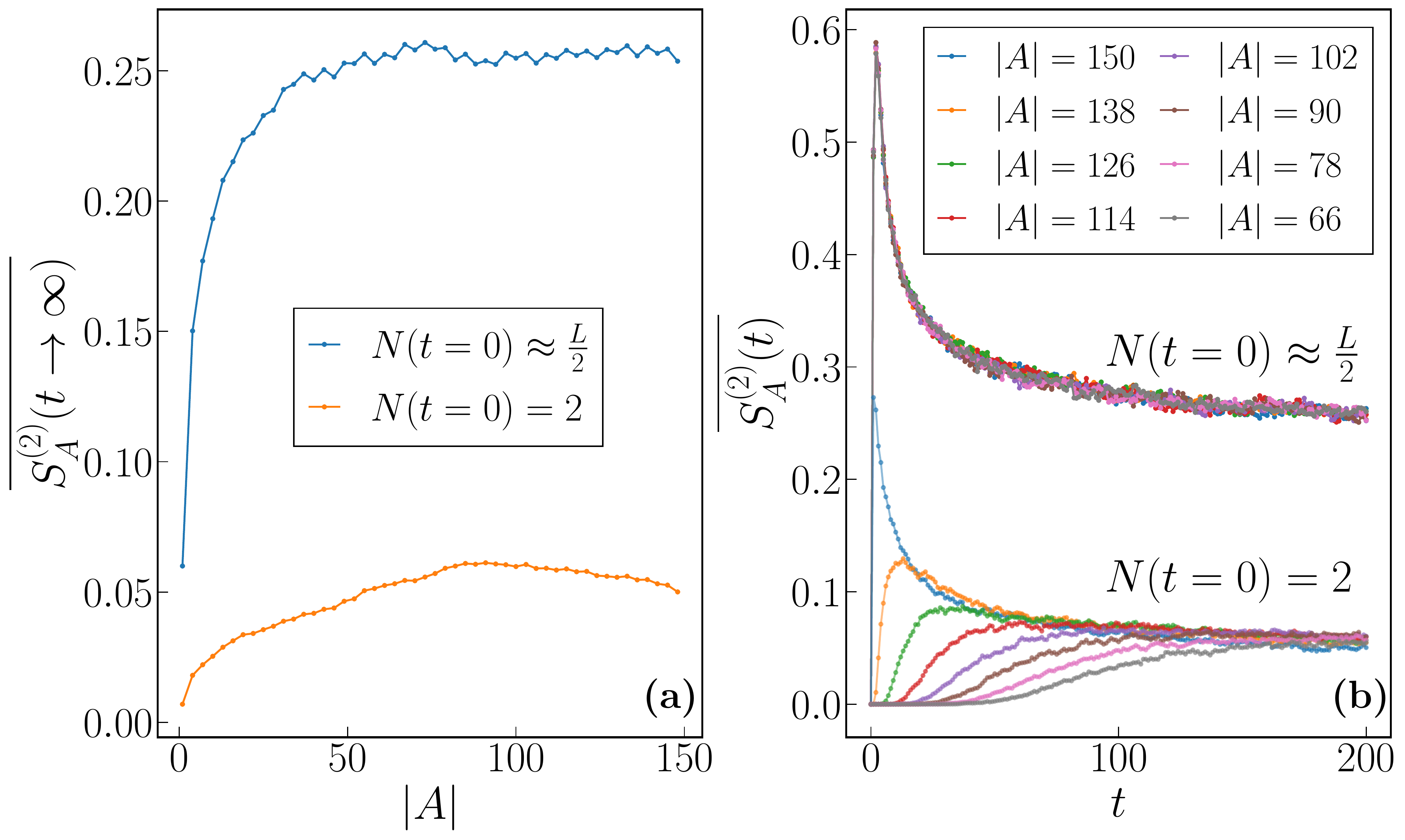}
\caption{(a) The steady state entanglement entropy remains independent of subsystem size, regardless of the initial state. (b) The difference in the early time growth visualized across different initial conditions and subsystem sizes. In both figures, $p=0.45 < p^n_c$ and $r=1$.}
\label{fig:2LM_EE} 
\end{figure}

The computational resources required to simulate the system increase exponentially as one nears the critical point $p_c^{EE}$ of the MIPT from above. Nonetheless, we provide strong evidence that these two transitions happen at \textit{different} points in our system. As we further lower $p$ such that $p<p_c^n\approx 0.55$, we find that the system remains in the area-law entangled phase despite a finite domain wall density, as is demonstrated by the scaling of $\overline{S^{(2)}_A}$ with subsystem size in Fig.~\ref{fig:2LM_EE}(a). In SM, we provide a heuristic argument based on the correlation length to explain why these two transitions in general should be different. We thus conclude that the critical point for the MIPT $p_c^{EE} < p_c^n$. This is also consistent with the general expectation that the entanglement transition for the individual trajectories must precede that for the quantum channel, if there is one.

\textit{Discussion.-} The physics discussed in this work remains unchanged if the 3-qubit unitary gates are replaced by $k$-qubit unitary gates ($k>3$) which leave $\underbrace{|00\ldots 0\rangle}_k$ and $\underbrace{|11\ldots 1\rangle}_k$ invariant (up to a $U(1)$ phase) and act as Haar random unitaries within the remaining $(2^{k}-2)$-dimensional subspace. When $k=2$, however, the local unitary gates coincide with those for a $U(1)$-symmetric Haar random circuit. As a result, the particles can only spread out diffusively rather than ballistically and hence cannot compete with any non-zero rate of particle annihilations induced by feedback. $\overline{n(t)}$ always decays as $t^{-1/2}$ for any $q>0$; there is \textit{no} MIPT and individual trajectories are immediately driven to an area-law entangled phase~\cite{SM}. This is in sharp contrast to $U(1)$-symmetric hybrid circuits without feedback, where an MIPT is present~\cite{Agrawal_2022}.

In this work, we have argued that the two transitions -- MIPTs and order-disorder transitions -- are generally unrelated. As we increase $r$, the difference between $p_c^{EE}$ and $p_c^n$ becomes smaller but remains finite when $r=1$. In SM, we consider models in which the essential physics is unchanged, but the difference between these two transitions is more pronounced. This is in contrast to a free fermion system subject to non-unitary dynamics with feedback that was recently studied in Ref.~\cite{buchhold2022revealing}. It remains to be seen if models with interactions can exhibit these transitions at the same point, while a broader understanding of the conditions that can facilitate this coincidence of these transitions is desirable.

We conclude by commenting on the experimental realizability of the order-disorder transition, spurred by the recent implementation of adaptive quantum circuits with high fidelity \cite{foss2023experimental,iqbal2023topological}. Overhead can be drastically reduced by estimating $\overline{n(t)}$ from measurement outcomes obtained over the course of the circuit evolution, instead of preparing the states at different times and then separately performing measurements. Furthermore, we numerically find that this transition is robust to imperfect rotations in the feedback and decoherence, but susceptible to other types of noise~\cite{SM}.

{\it Note added.-} Shortly after our paper appeared on the arXiv, we became aware of an independent work where similar results were obtained in a different setup~\cite{o2022entanglement}.

{\it Acknowledgements.-} We thank Sebastian Diehl for explaining the results of Ref.~\cite{buchhold2022revealing}, and Ehud Altman for useful discussions. This research is supported in part by the Google Research Scholar Program and is supported in part by the National Science Foundation under Grant No. DMR-2219735. Z.-C.Y. is supported by a startup fund at Peking University. We gratefully acknowledge computing resources from Research Services at Boston College and the assistance provided by Wei Qiu. 

\nocite{*}
\bibliography{reference}
\end{document}


\title{Supplemental Material: Entanglement Steering in Adaptive Circuits with Feedback }

\author{Vikram Ravindranath}
\affiliation{Department of Physics, Boston College, Chestnut Hill, MA 02467, USA}

\author{Yiqiu Han}
\affiliation{Department of Physics, Boston College, Chestnut Hill, MA 02467, USA}

\author{Zhi-Cheng Yang}
\email{zcyang19@pku.edu.cn}
\affiliation{School of Physics, Peking University, Beijing 100871, China}
\affiliation{Center for High Energy Physics, Peking University, Beijing 100871, China}

\author{Xiao Chen}
\affiliation{Department of Physics, Boston College, Chestnut Hill, MA 02467, USA}

\onecolumngrid
\setcounter{equation}{0}
\setcounter{figure}{0}
\setcounter{table}{0}
\setcounter{page}{1}
\makeatletter
\renewcommand{\theequation}{S\arabic{equation}}
\renewcommand{\thefigure}{S\arabic{figure}}
\renewcommand{\bibnumfmt}[1]{[S#1]}
\renewcommand{\citenumfont}[1]{S#1}

\maketitle
\section{Particle density transition of the classical bit-string dynamics}

Inspired by the quantum dynamics under the hybrid circuit with feedback, we proposed a classical stochastic process that traces the dynamics of the bit strings in the quantum state shown in Fig.~3. Under the 3-site unitary gates, the bit string either stays invariant if there are no domain walls or is mapped to another bit string within the ensemble $\{|001\rangle,|010\rangle,|100\rangle,|011\rangle,|101\rangle,|110\rangle\}$ with equal probability. On the other hand, when the measured sites have opposite spins, the Pauli-$X$ operator rotates either spin and kills a pair of domain-walls (a pair of neighboring particles) with probability $r$. Since the measurement rate is $p$, this leads to an effective annihilation rate $q\equiv pr$. Therefore the domain-wall particles either diffuse, branch or annihilate in pairs. This is often referred to as branching-annihilating random walks (BAW) with even offspring,

\begin{equation}
   W \leftrightarrow 3W, \quad W + W \xrightarrow[]{q} \emptyset.
\end{equation}
BAW models with even offspring usually experience an absorbing phase transition that belongs to parity-conserving (PC) universality class, which is characterized by the additional symmetry that preserves the parity of the number of particles in the system. 

In this appendix, through investigating the domain-wall particle density transition of three different circuit setups, $\emph{i.e.}$, hybrid circuits with one/two/three sets of measurements with probability $q$ per time step, we will show that our classical bit-string model belongs to PC universality class. We consider two initial conditions which lead to different scaling behaviors in the particle density under the same dynamics: (a) the seeding process beginning with a pair of adjacent domain-wall particles, and (b) the purification process beginning with a randomly occupied state ($\emph{i.e.}$, a random bit-string). 

\begin{figure*}
  \centering
  \subfigure[]{
    \includegraphics[width=0.33\textwidth]{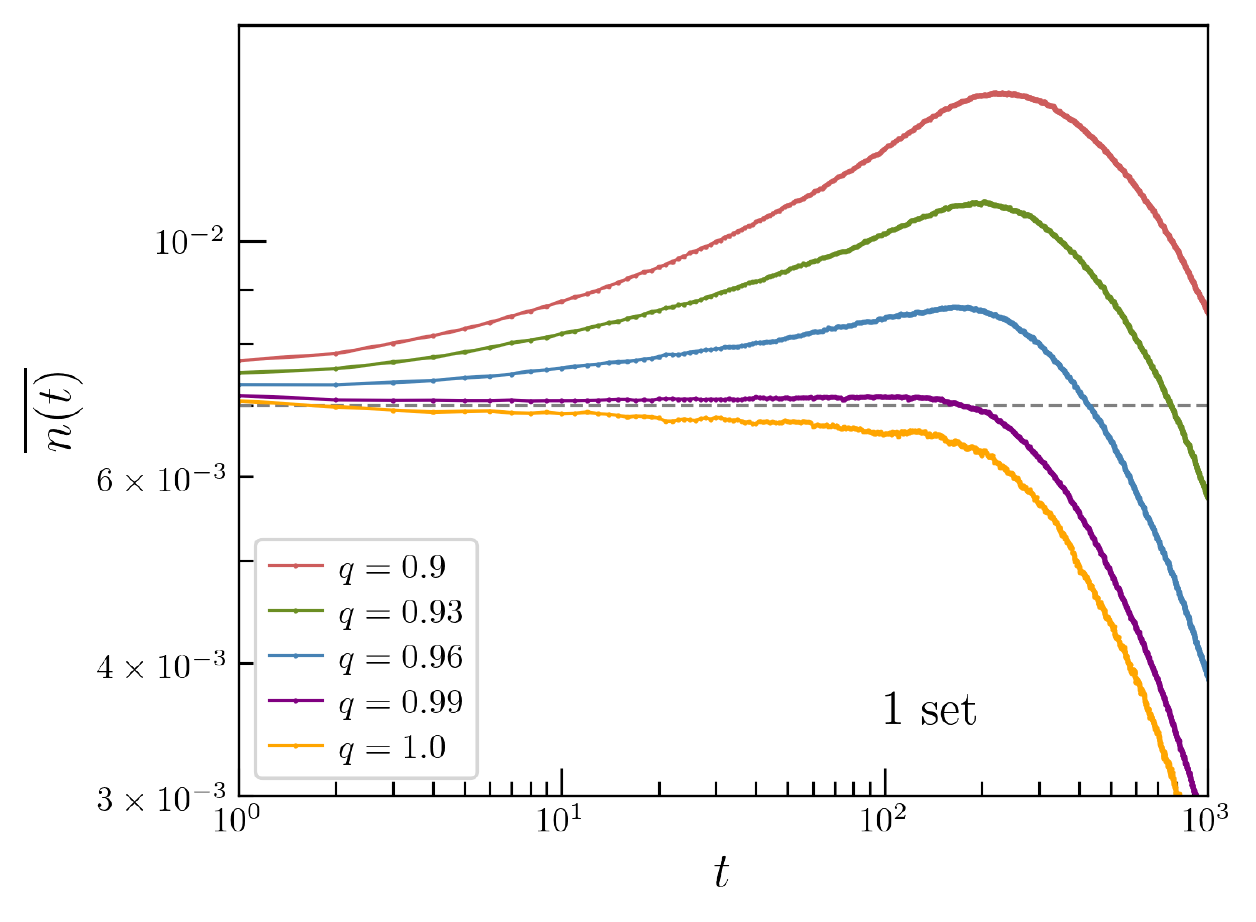}
    \label{fig:1S}
  }%
  \subfigure[]{
    \includegraphics[width=0.33\textwidth]{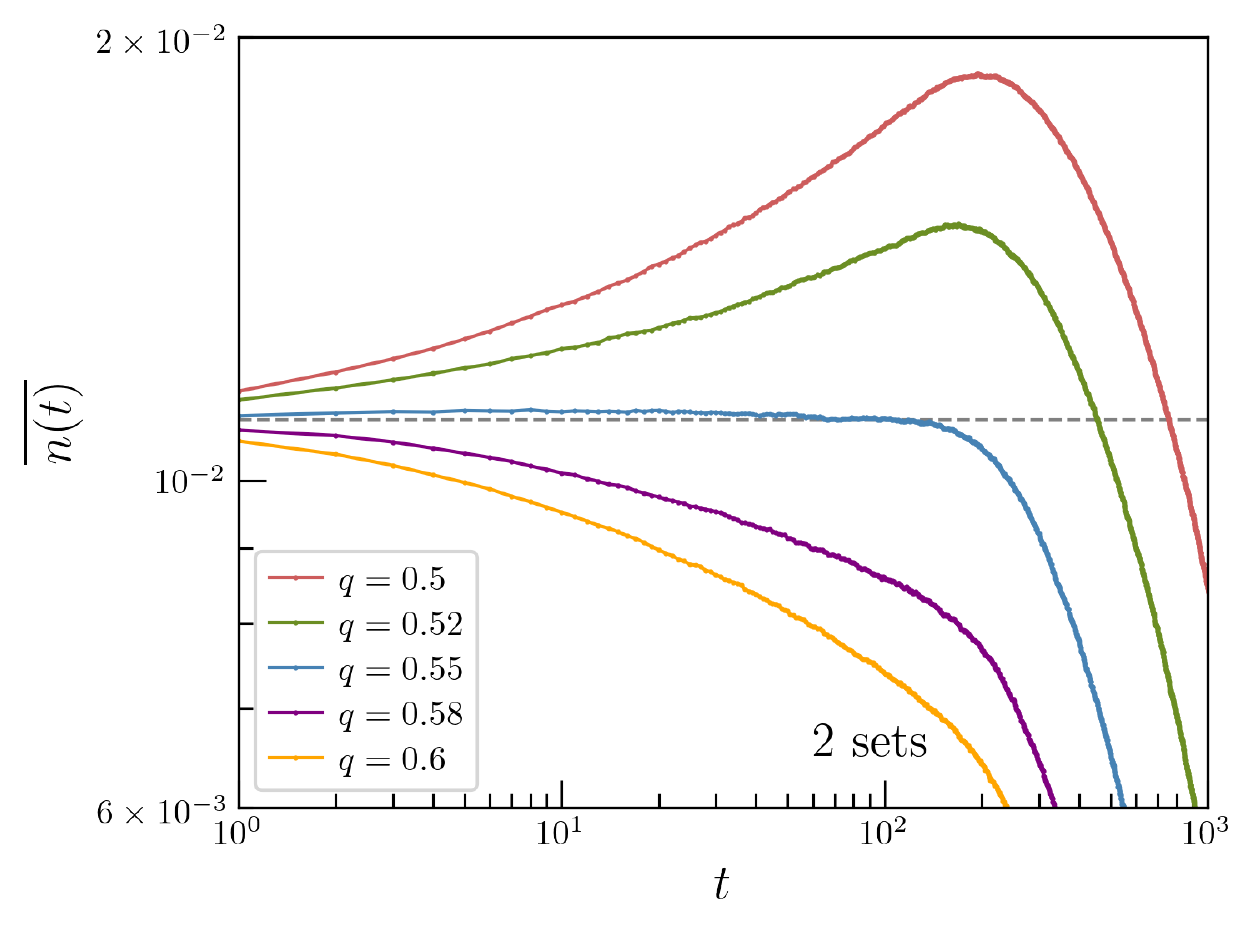}
    \label{fig:2S}
  }%
  \subfigure[]{
    \includegraphics[width=0.33\textwidth]{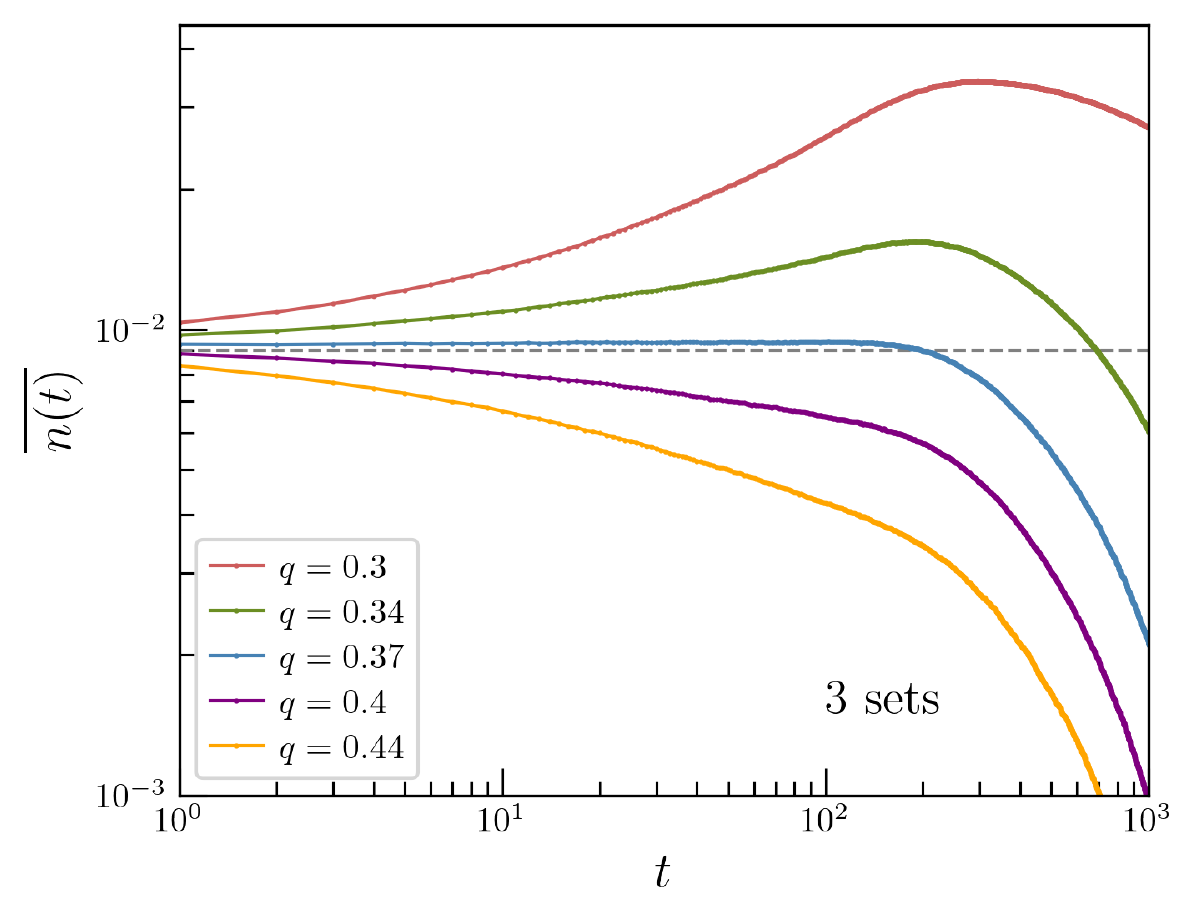}
    \label{fig:3S}
  }
  
  \subfigure[]{
    \includegraphics[width=0.33\textwidth]{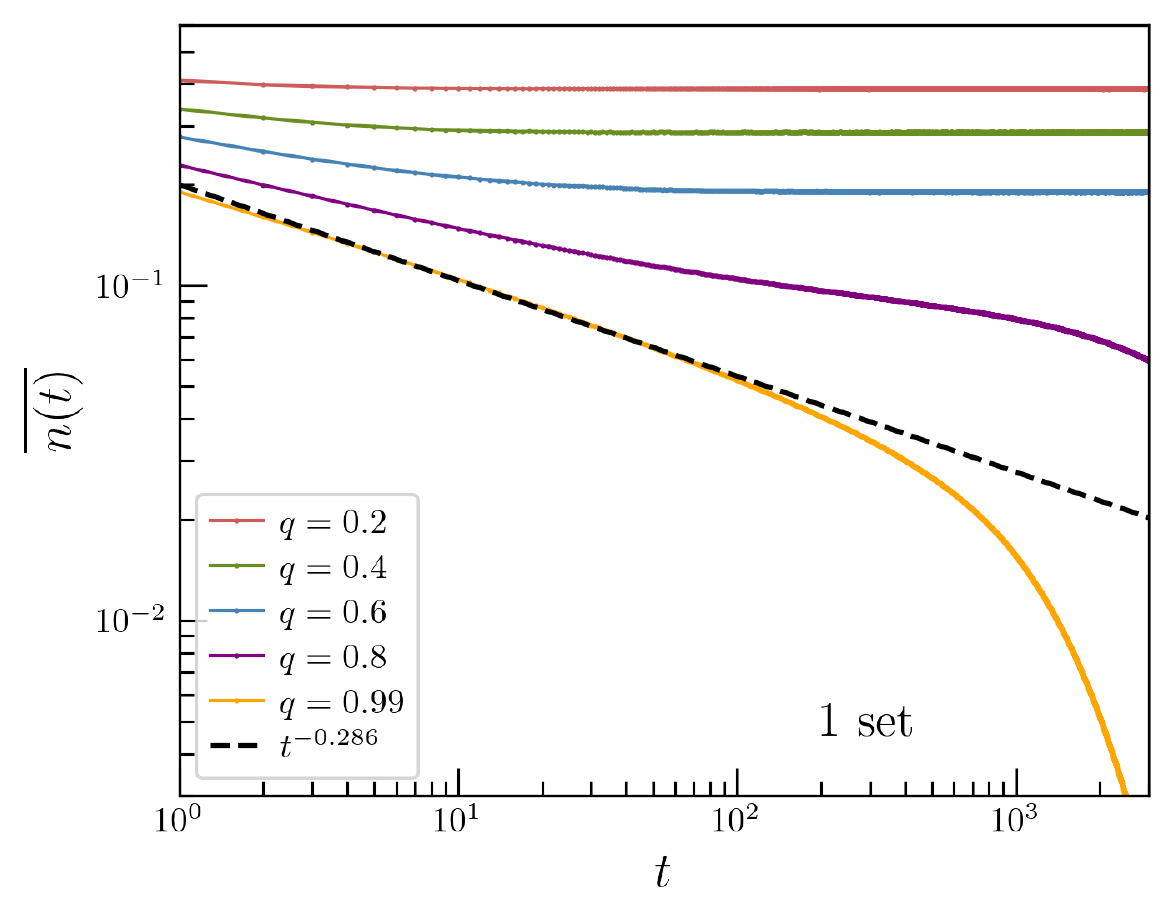}
    \label{fig:1P}
  }%
  \subfigure[]{
    \includegraphics[width=0.33\textwidth]{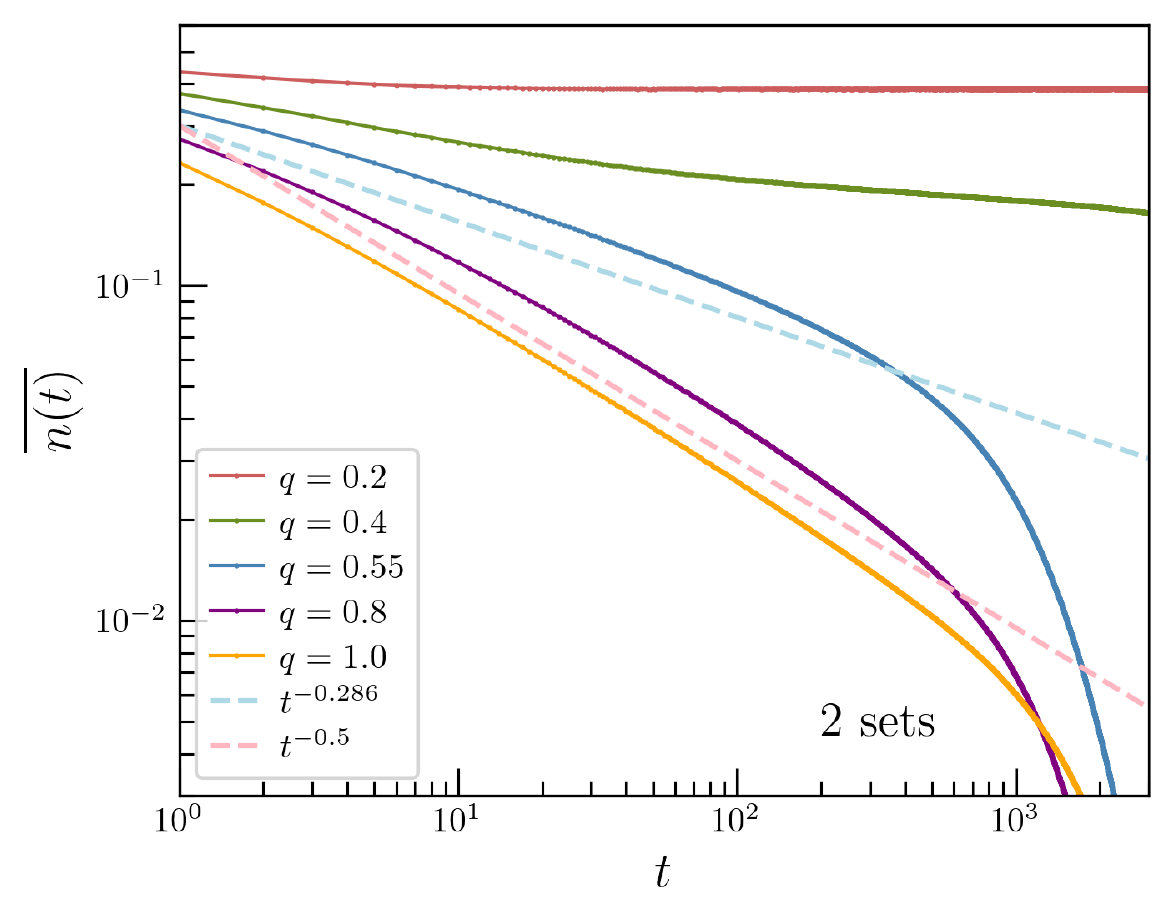}
    \label{fig:2P}
  }%
  \subfigure[]{
    \includegraphics[width=0.33\textwidth]{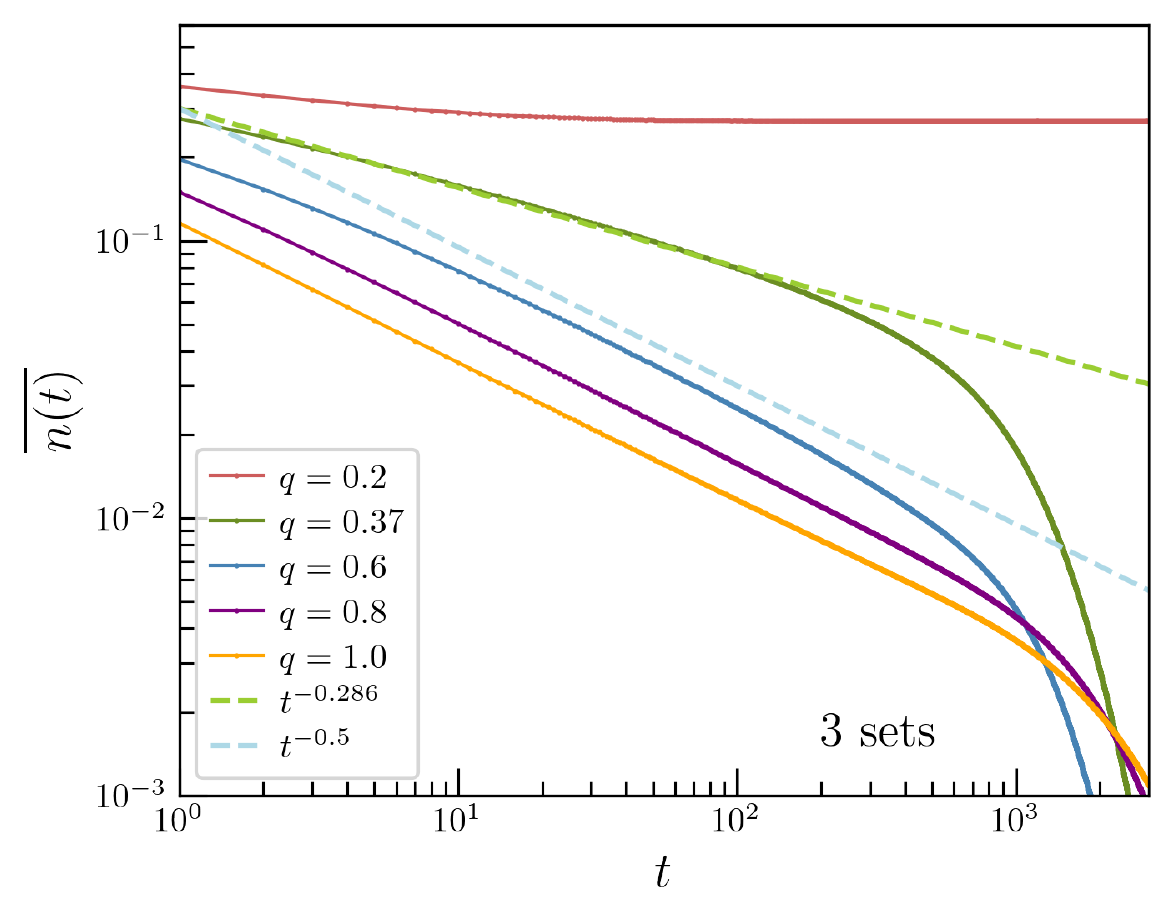}
    \label{fig:3P}
  }
    
  \caption{We simulate the domain-wall particle density $\overline{n(t)}$ of the bit-string model with one set of measurements in (a) and (c), two sets of measurements in (b) and (e), and three sets of measurements in (c) and (f). Among them, (a), (b), and (c) are the results of the seeding process starting with a pair of adjacent particles, (d), (e), and (f) are the results of the purification process starting with a random state. All of the data are computed for system size $L=300$ under open boundary condition (OBC) over a variety of measurement rates $q$, and we plot $\overline{n(t)}$ vs $t$ on a log-log scale.}
  \label{fig:PC_transition}
\end{figure*}

The numerical results are shown in Fig. \ref{fig:PC_transition}. We vary the annihilation rate $q$ and calculate the scaling behavior of the mean domain-wall particle density $\overline{n(t)}\equiv \overline{N(t)}/L$, where $\overline{N(t)}$ is the mean domain-wall particle number and $L$ is the system size. We find that except for the model with one layer of measurements per time step, the rest of the setups all experience a phase transition while adjusting $q$. For the purification process beginning with a random bit string state, when $q<q_c$, the system stays at an active steady state with a finite particle density. When $q=q_c$, $\overline{n(t)}\sim t^{-\theta}$ with $\theta=0.286$. When $q>q_c$, the particles perform annihilation-dominated BAW and the particle density decays diffusively, i.e., $\overline{n(t)}\sim t^{-1/2}$. For the seeding process starting with a pair of adjacent domain-wall particles, despite the finite-size effect, when $q<q_c$, the system approaches an active state. At $q=q_c$, $\overline{n(t)}\sim t^{\delta}$ with $\delta=0$. When $q>q_c$, $\overline{n(t)}$ still decreases diffusively (not shown in the plot). These exponents are universal and agree with the numerical findings of the PC universality class that $\theta=0.286$, $\delta=0$ for $q=q_c$, and $z=2$ for $q>q_c$.

\begin{figure*}[!]
  \centering
  \subfigure[]{
    \includegraphics[width=0.33\textwidth]{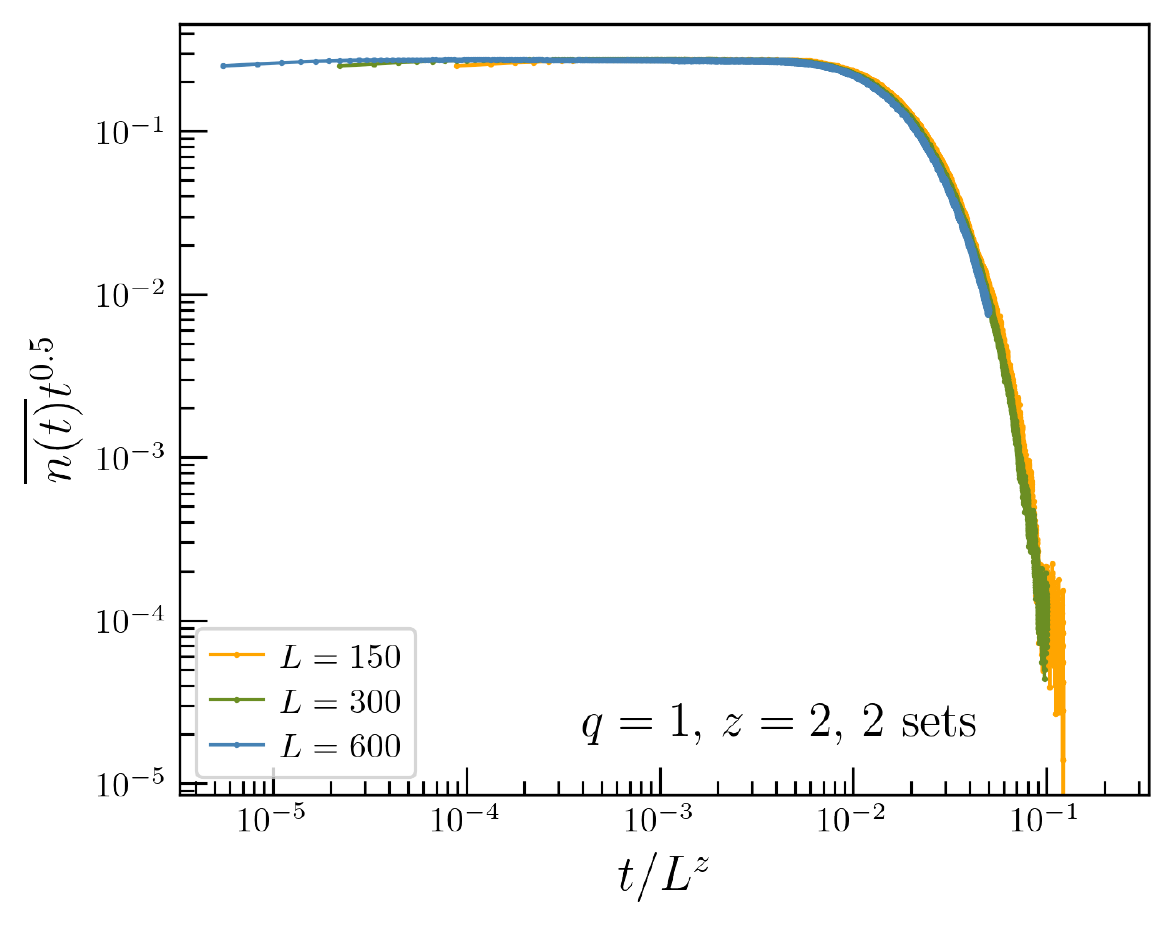}
    \label{fig:2P_dc_t_p_1}
  }%
  \subfigure[]{
    \includegraphics[width=0.33\textwidth]{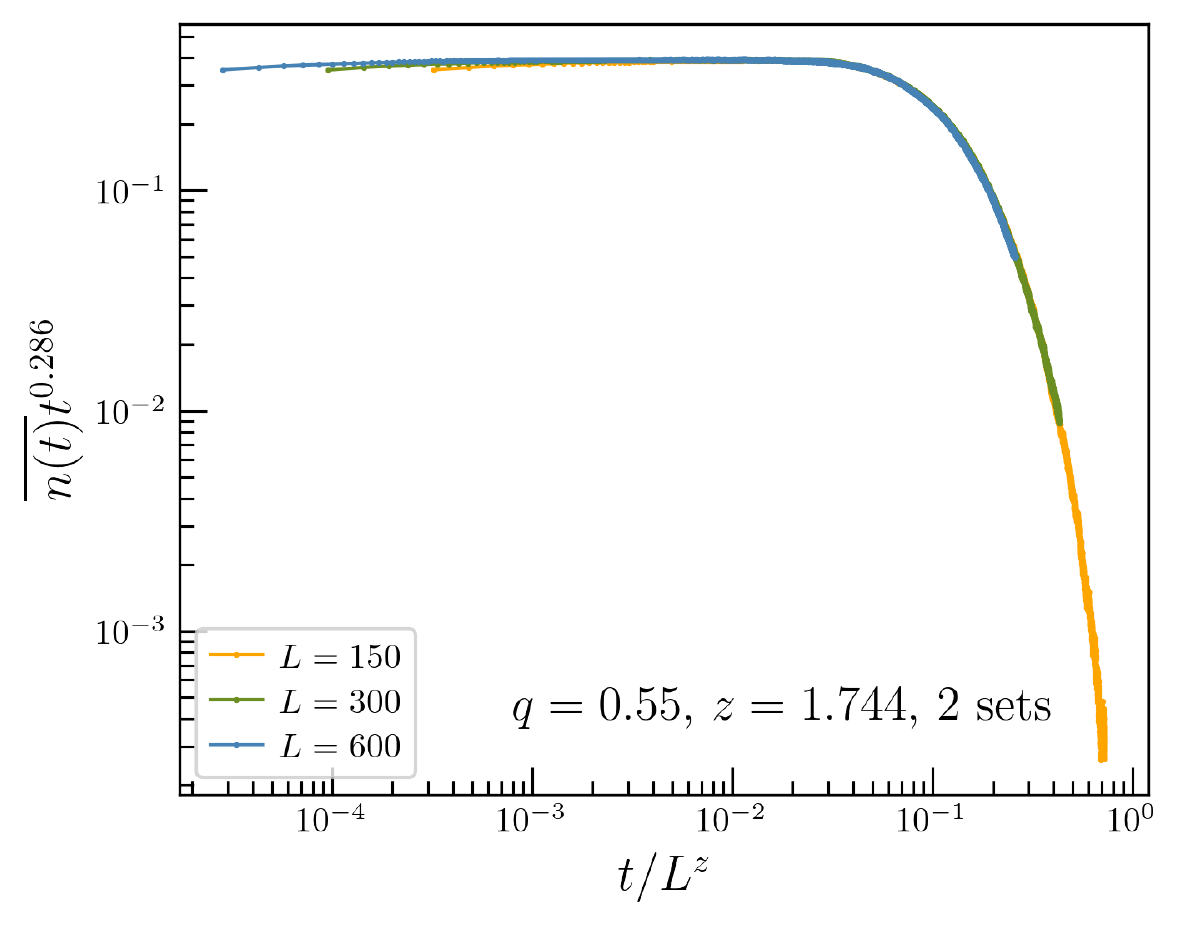}
    \label{fig:2P_dc_t}
  }%
  \subfigure[]{
    \includegraphics[width=0.33\textwidth]{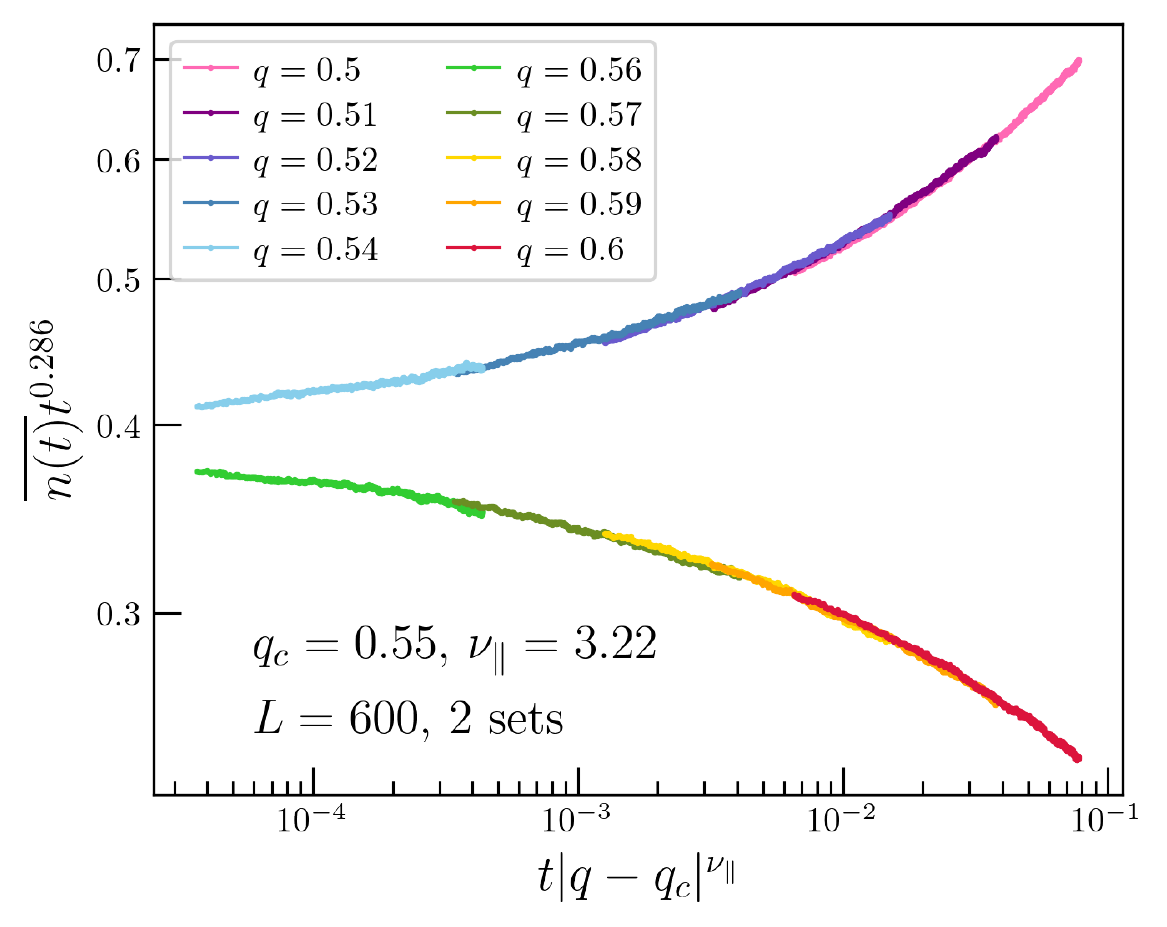}
    \label{fig:2P_dc_p}
  }
   \caption{Finite-size data collapse of the domain-wall particle density $\overline{n(t)}$ of the classical bit-string model with two sets of measurements for different system sizes at (a) $q=1$, and (b) $q=q_c=0.55$, respectively. (c) Data collapse using the scaling form Eq.~(\ref{eq:scaling}), for $|q-q_c|=0.01, 0.02, \dots, 0.05$ at $L=600$ for $t\in[100,1200]$. We use periodic boundary condition (PBC) for numerical simulations. }
\end{figure*}

The two different initial conditions also help us to pin down the critical point. We do find that for $L=300$, the bit-string model with one layer of measurements has a critical point $q_c=0.99$. However, this is very close to $q=1$ which is probably due to the finite size effect. Hence, we claim that the case with only one set of measurements per time step doesn't have a phase transition. For models with two and three sets of measurements per time step, the critical points are $q_c=0.55$ and $q_c=0.37$ respectively, which agree with the corresponding  $q_c$'s found via direct simulations of the circuit evolution.

In addition, we have performed a detailed finite-size scaling analysis of the domain wall density of the purification process in the vicinity of the critical point. We expect that the domain wall density satisfies the following scaling form:

\begin{equation}
    n(t,q)=t^{-\theta}f(t/L^z,|q-q_c|t^{1/\nu_\|}),
\label{eq:scaling}
\end{equation}
where $z^{PC}=1.744$ at the critical point $q_c=0.55$ and $\nu_{\|}^{PC}=3.22$ \cite{zhong1995universality}. As shown in Fig.\ref{fig:2P_dc_t} and \ref{fig:2P_dc_p}, we find excellent agreement with the above scaling form using known exponents of the PC universality class. We have also examined the data collapse in the critical phase where $z^{PC}=2$ for $q>q_c$ in Fig.\ref{fig:2P_dc_t_p_1}. This further confirms that our model belongs to the PC universality class.

\begin{figure}[]
\includegraphics[width=.45\textwidth]{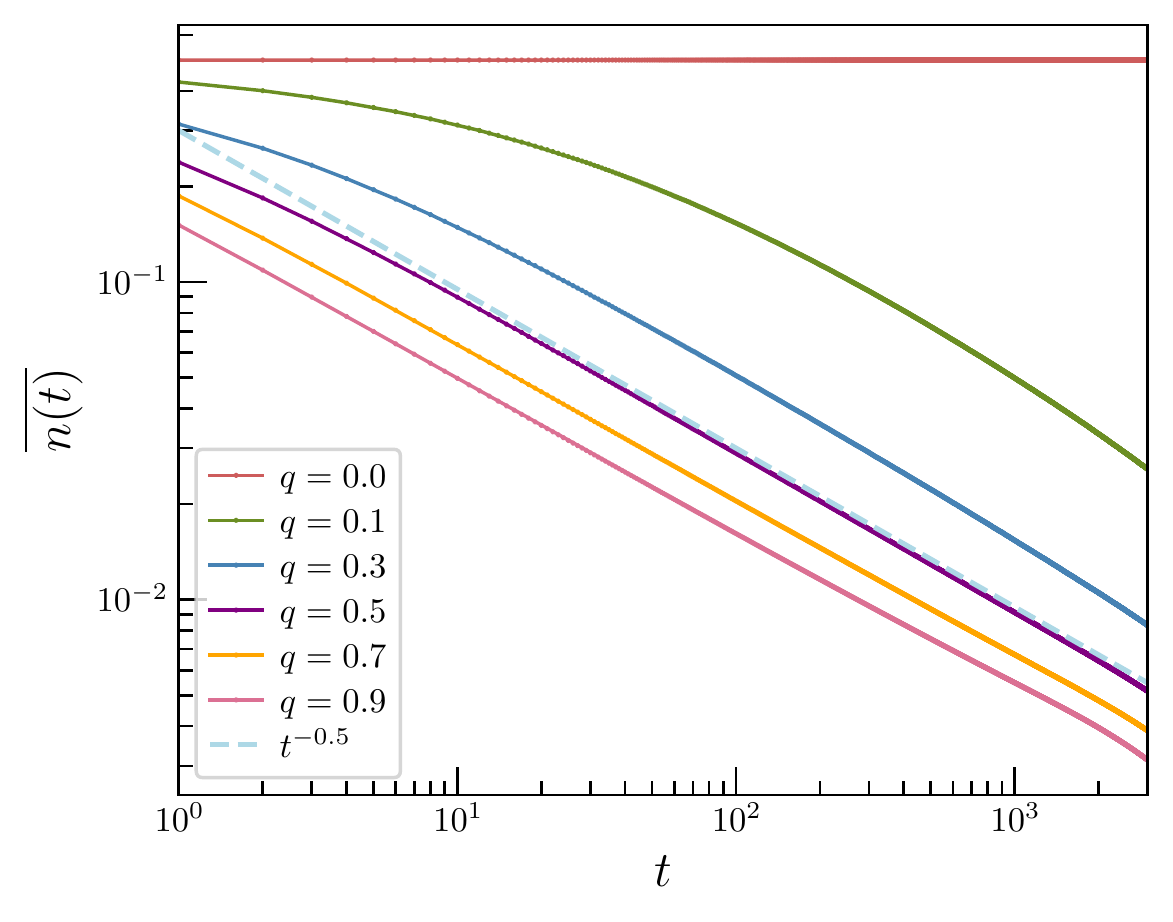}
\caption{The particle density $\overline{n(t)}$ vs $t$ plotted on a log-log scale for the purification process under the circuit with two sets of 2-site unitary gates and two sets of measurements with probability $q$ per unit time step. When $q=0$, $\overline{n(t)}$ does not decay with time. Once $q>0$, the particle density decays as $\overline{n(t)}\sim t^{-1/2}$. All of the data are collected for system size $L=300$ under OBC.}
\label{fig:k=2 bit string} 
\end{figure}

Finally, we consider replacing the 3-site unitary gates with 2-site unitary gates, which keep the bit strings $\{|11\rangle,|00\rangle\}$ unchanged and map the bit strings with domain walls randomly within the ensemble $\{|10\rangle,|01\rangle\}$. The $U(1)$ symmetry imposed by the 2-site unitary gates preserves the total spin and the domain-wall particles propagate diffusively, which is too weak to compete with the particle annihilation due to measurements. Therefore, there is no particle density phase transition and we have $\overline{n(t)}\sim t^{-1/2}$ for any $q>0$. This is verified by Fig.\ref{fig:k=2 bit string} in which we simulate $\overline{n(t)}$ of the purification process with two sets of 2-site unitaries interspersed with two sets of measurements per unit time.

\section{Results with 1 set of measurements}
 In this section, we present results for the set-up where the feedback scheme is implemented with only 1 set of measurements per time step, \textit{after} all 3 layers of unitary gates have been applied. As in the classical model, the transition in $n$ appears to occur extremely close to $pr=1$, rendering the observation of the ordered phase unfeasible, owing to finite-size effects. The transition points $r_c$ and $p^n_c$ are estimated as in the main text, by identifying the values of $r$ and $p$ at which $\overline{n(t)}$ does not change with time, beginning from an initial state with 2 adjacent domain walls in the center of the chain. It is found that when $p=1$, $r_c\approx0.99$, and $p_c^n\approx0.99$ when $r=1$.
 
 \begin{figure}[!t]
     \subfigure[]{\includegraphics[width=0.4\textwidth]{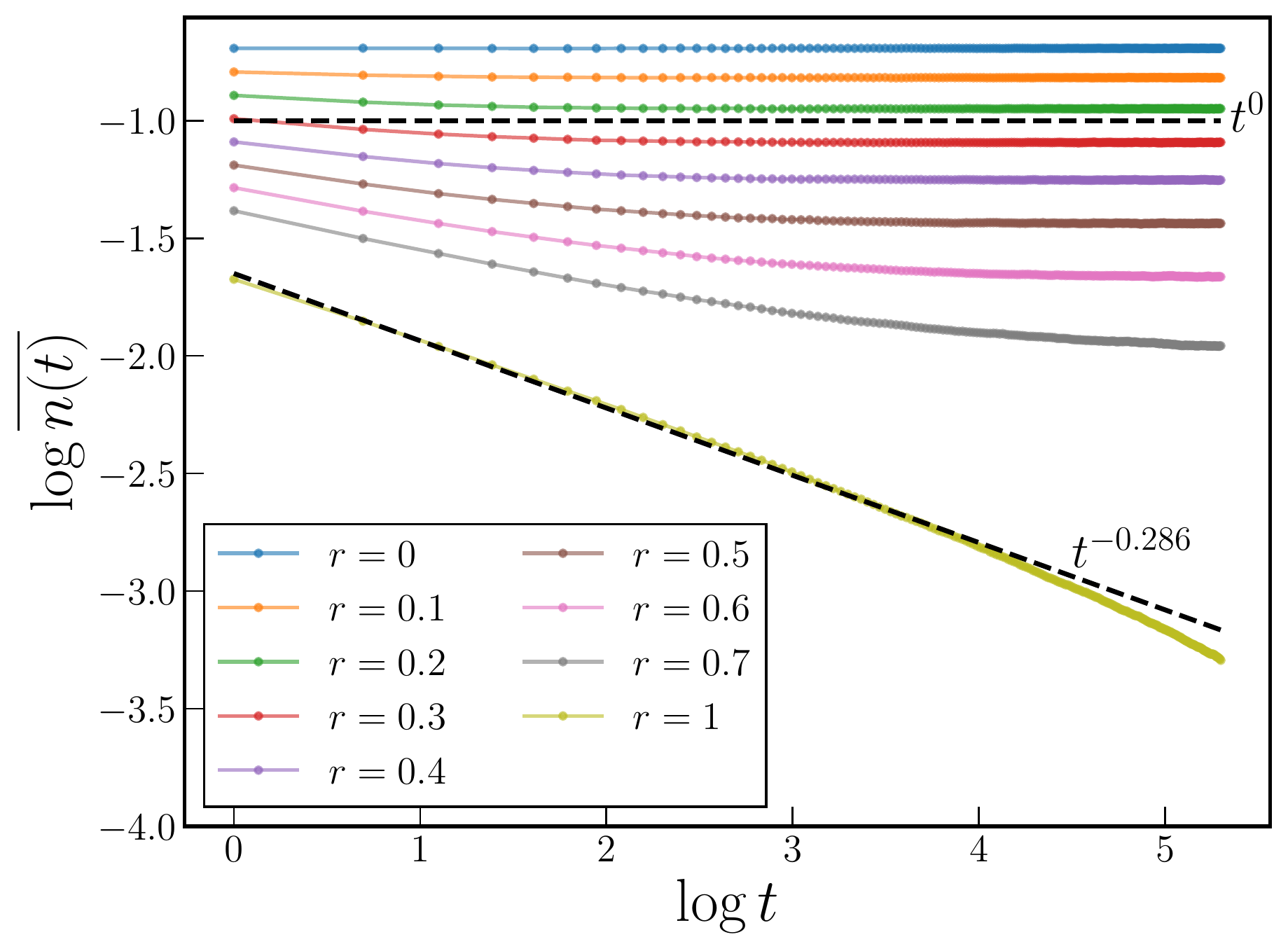}}\subfigure[]{\includegraphics[width=0.525\textwidth]{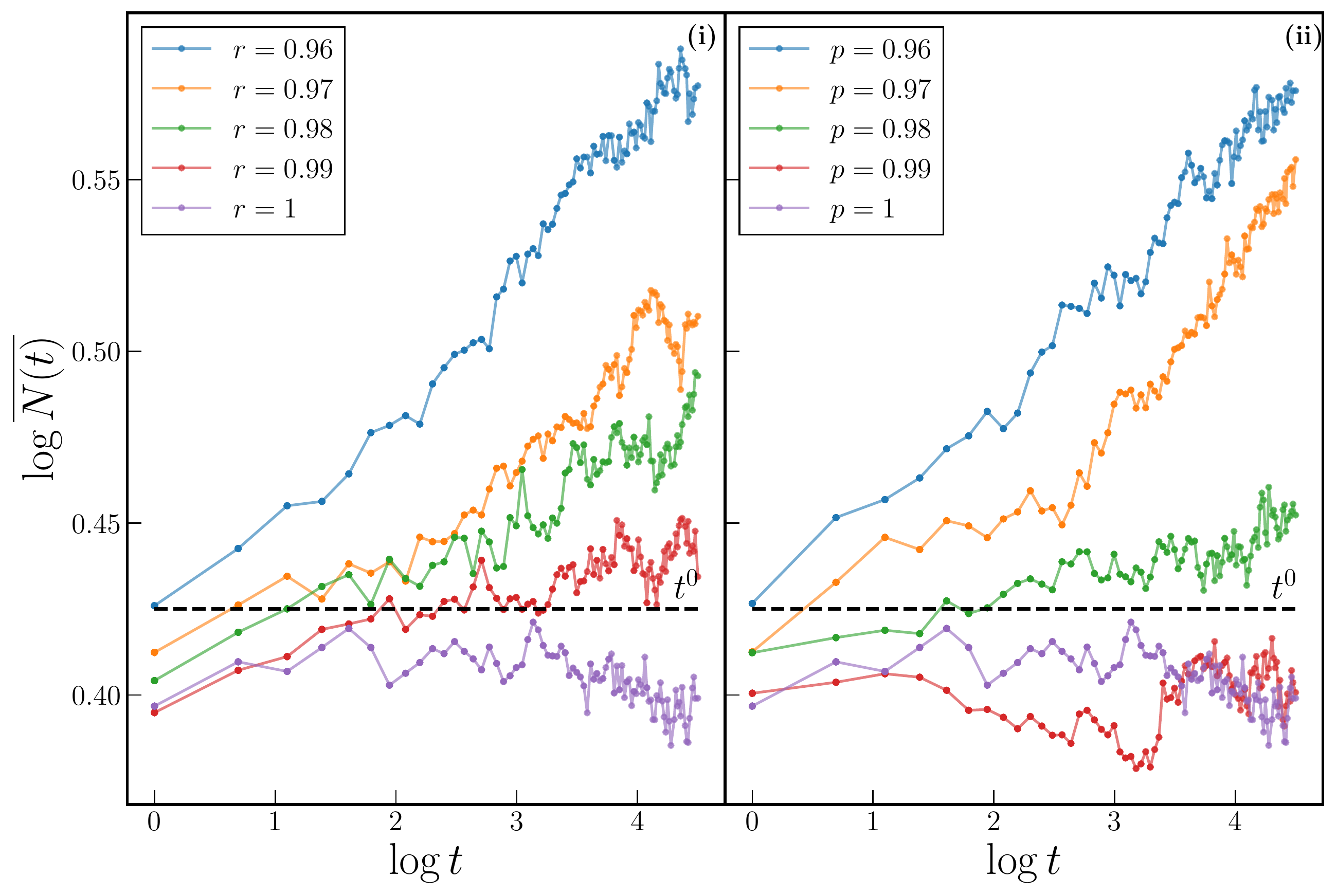}}
     \subfigure[]{\includegraphics[width=0.4\textwidth]{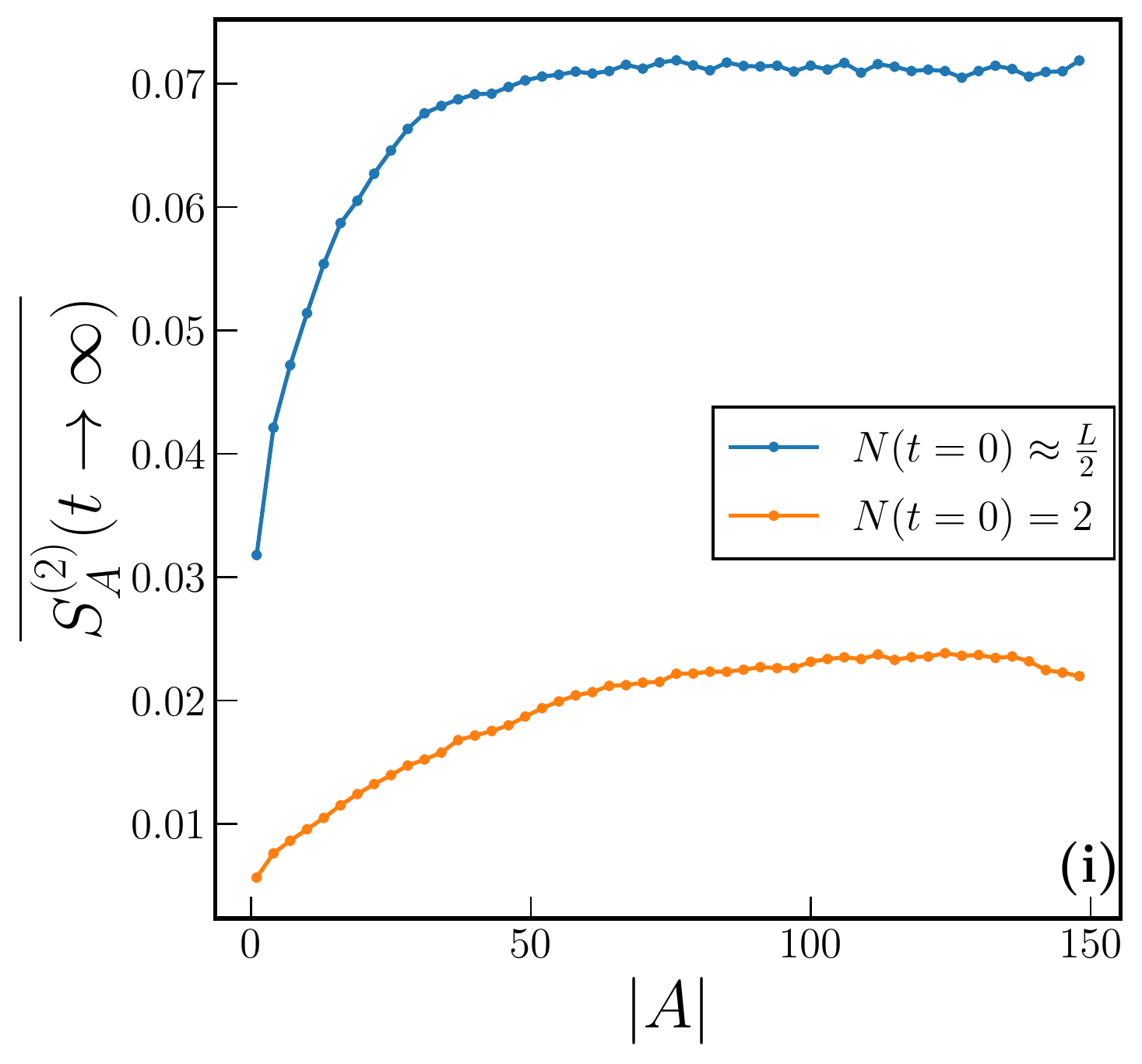}}\subfigure[]{\includegraphics[width=0.5\textwidth]{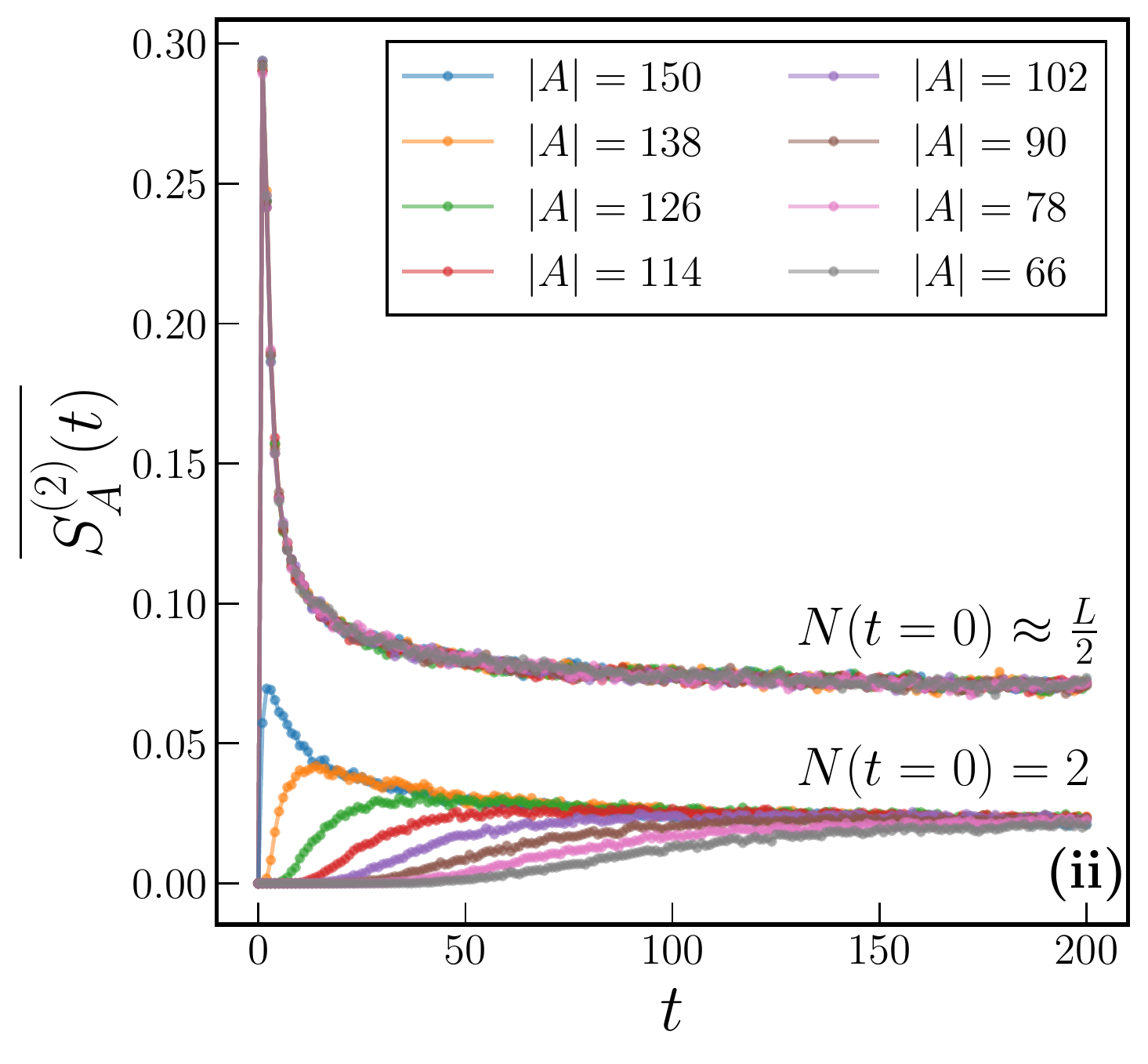}}
     \caption{Numerically obtained results for the quantum model with 1 set of measurements after 3 layers of unitary gates, for $L=300$, and results averaged over $2\times10^4$ realizations. (a) The domain wall density $n$ at $p=1$, showing a critical scaling of $\overline{n(t)}\approx t^{-0.286}$ near $r\approx1$. The steady state domain wall density appears finite for $r\lesssim0.5$, but finite size effects prevail for $r\gtrsim0.7$. (b) The number of domain walls, beginning from an initial state with two adjacent domain walls in the center of the system. The data affords an estimation of $q^n_c \approx0.99$. (i) $p=1$ and (ii) $r=1$. (c) The steady state entanglement scaling for $r=1, p=0.8 < p^n_c$. The entanglement satisfies area law scaling and  shows a slight dependence on the initial condition. (d) The entanglement dynamics for two different initial states for $r=1, p=0.8 < p^n_c$. For $N(t=0)=2$, the two domain walls are located close to the center.  
     }
     \label{fig:1LM}
 \end{figure}
 
 We find that $p^{EE}_c < p^n_c$ in this case as well. The characteristic dependence of the entanglement growth on the initial state is also present here, since the entangling unitary gates act trivially on large parts of the system when considering initial conditions with a vanishing \textit{density} of domain walls. These results provide compelling evidence that (a) the two transitions are indeed different and (b) The qualitative features of this family of adaptive quantum circuits are largely insensitive to the microscopic details of the circuit. These conclusions strengthen, and are strengthened by, the connection to the classical dynamics.

\section{Effects of Imperfect Circuits}

We were initially motivated by strategies that could facilitate the realization of the measurement-induced phase transition in present-day NISQ-era quantum devices. In this spirit, we study the effects that noise might have on the different dynamical phenomena we observe. In particular, we consider (i) the effects of imperfect corrective rotations and (ii) the role of decoherence.

\subsection{Imperfect Corrective Rotations}
\begin{figure}
    \centering
    \subfigure[]{\includegraphics[width=0.45\textwidth]{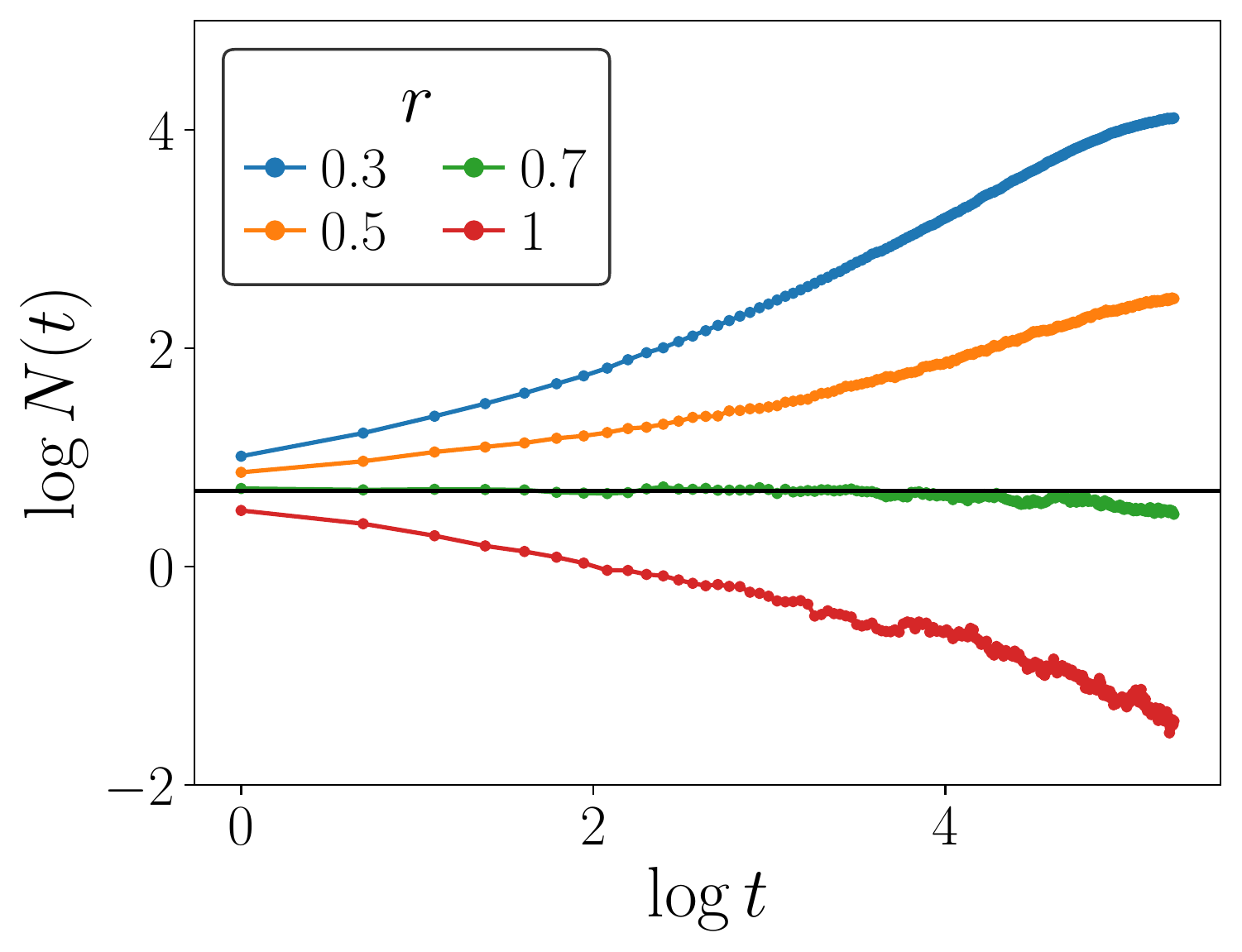}} \subfigure[]{\includegraphics[width=0.45\textwidth]{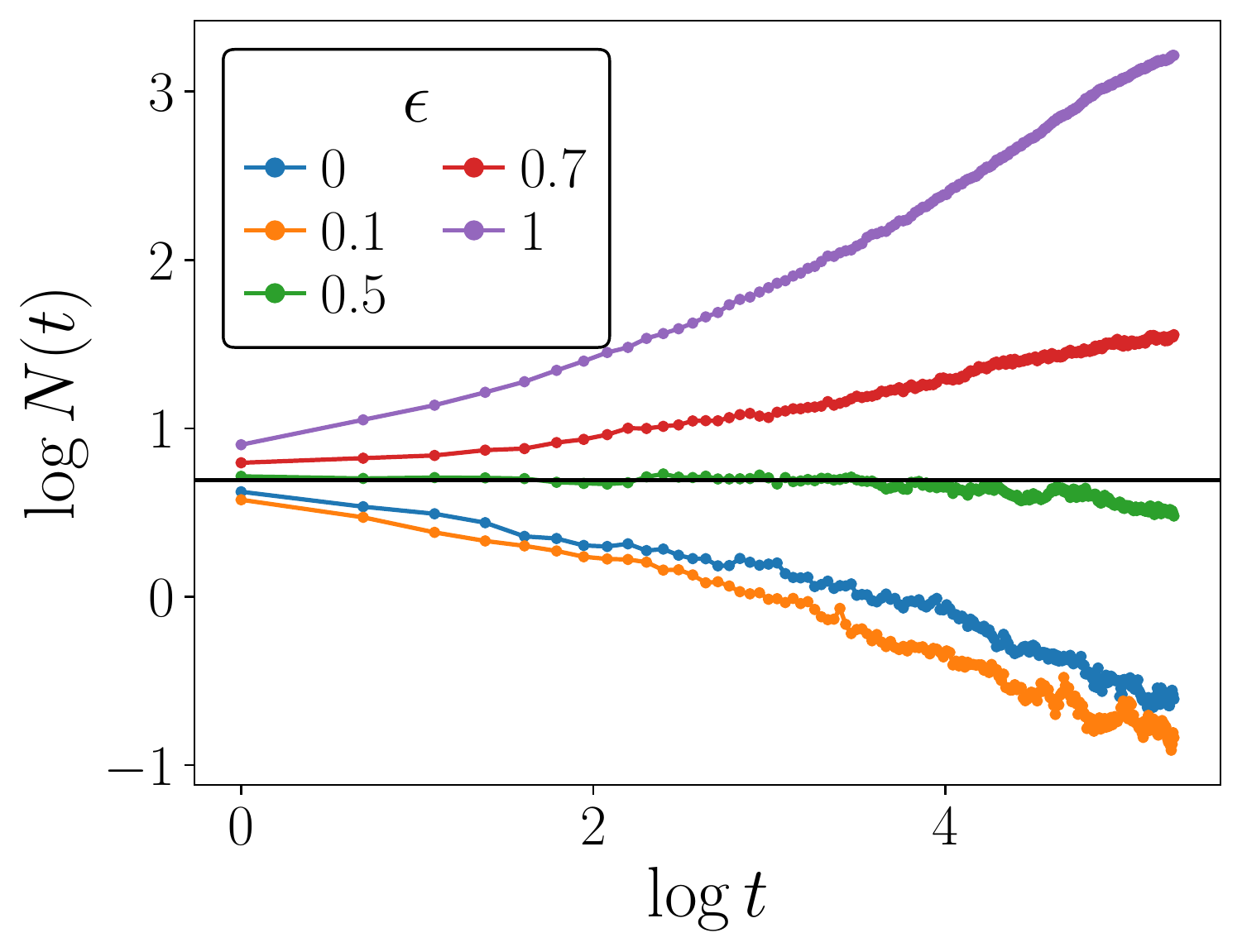}}
    \caption{The effects of imperfect rotations on the domain wall density, shown only to have the effect of renormalizing $r$ as given in \cref{eq:corrective_final}. Simulations are performed on $L=300$ spins, starting with a pair of adjacent domain walls in the center of the chain, and averaged over $2\times 10^3$ realizations. $r_c\approx0.55$ in the absence of noise. (a) $r$ is varied, holding $\epsilon=0.5$ fixed. The critical $r$ shifts downward to $r_c\qty(\epsilon=0.55)\approx 0.7$. (b) With $r$ fixed at $r=0.7$, a transition between the two dynamical phases is driven by tuning the strength of the noise $\epsilon$.}
    \label{fig:err_Rot}
\end{figure}

Our set-up involves measuring the operator $Z_i Z_{i+1}$ on neighboring sites, which can result in the outcomes $\pm1$. If the outcome $-1$ is observed, then, with probability $r$, either the qubit at location $i$ or at $i+1$ is rotated by $\pi$ about the X-axis. We account for imperfections in this process by considering noisy, imperfect rotations of the form

\begin{align}
    R_{\epsilon,i,t} &= e^{-i\theta_{\epsilon,i,t} X}\nonumber\\
    \theta_{\epsilon,i,t} &\equiv \frac{\pi}{2} \left(  1 + \epsilon \tilde{x}_{i,t}\right)
\end{align}
where $\tilde{x}$ is a random number independently drawn from $[-1,1]$ at each time $t$ and site $i$ when a corrective rotation is to be applied. $\epsilon$ dictates the strength of the noise. The results are presented in \cref{fig:err_Rot}.

The transition still persists even in the presence of noise. We see that the effect of the noise in the rotation angle is to merely renormalize the rate $r$ at which errors are corrected. The renormalized value can be explicitly obtained by considering the bit-string picture. We fix $p=1$ so that the feedback rate is solely controlled by $r$.
In the absence of any noise $(\epsilon = 0)$, upon finding an outcome of $Z_i Z_{i+1} = -1$, there are two possible scenarios - 

\begin{equation}
    \expval{Z_i Z_{i+1}}_{a.m.} = \begin{cases}
        +1, &\text{ with probability } r\\
        -1, &\text {with probability } 1 - r
    \end{cases}
    \label{eq:corrective}
\end{equation}
($a.m.$ denotes ``after measurement"). When $\epsilon\neq0$, it will be helpful to write (with the $i,t$ labels suppressed) $R_\epsilon = \cos{\theta_\epsilon} + i X \sin{\theta_\epsilon}$. Assuming the correction occurs at site $i$, \cref{eq:corrective} is modified to give

\begin{equation}
    \expval{Z_i Z_{i+1}}_{a.m.} = \begin{cases}
        \expval{R^\dagger_\epsilon Z_i R_\epsilon Z_{i+1}}_{a.m.}, &\text{ with probability } r\\
        -1, &\text{ with probability } 1 - r
    \end{cases}
    \label{eq:corrective1}
\end{equation}

We can further expand $R^\dagger_\epsilon Z_i R_\epsilon Z_{i+1}$ as 
\begin{equation}
    \begin{aligned}
    \cos^2\qty(\theta_\epsilon) \expval{Z_i Z_{i+1}=-1}_{a.m.} &- \sin^2\qty(\theta_\epsilon) \expval{Z_i Z_{i+1}=+1}_{a.m.}\\
     = \sin^2 \qty(\frac{\pi \epsilon\tilde{x}}{2}) (-1) &+ \cos^2 \qty(\frac{\pi \epsilon\tilde{x}}{2}) (+1).
    \end{aligned}
    \label{eq:err_result}
\end{equation}

Since the measurements are frequent and we expect $Z_i Z_{i+1}$ to have a definite value, we can interpret the result of \cref{eq:err_result} as a stochastic process with the update rule

\begin{equation}
\begin{aligned}
    \expval{R^\dagger_\epsilon Z_i R_\epsilon Z_{i+1}}_{a.m.} = \begin{cases}
        -1, &\text{ with probability } \sin^2 \frac{\pi \epsilon\tilde{x}}{2}\\
        +1, &\text{ with probability } \cos^2 \frac{\pi \epsilon\tilde{x}}{2}
    \end{cases}
\end{aligned}
\label{eq:err_result1}
\end{equation}

Combining \cref{eq:corrective1,eq:err_result1}, we finally have

\begin{equation}
    \expval{Z_i Z_{i+1}}_{a.m.} = \begin{cases}
        +1, &\text{ with probability } r\cos^2 \qty(\frac{\pi \epsilon\tilde{x}}{2})\\
        -1, &\text {with probability } 1 - r + r\sin^2 \qty(\frac{\pi \epsilon\tilde{x}}{2}).
    \end{cases}
    \label{eq:corrective_final}
\end{equation}

By averaging over $\tilde{x}$, we arrive at an expression for the renormalized value of $r$

\begin{equation}
    r_{\text{renorm}} = \frac{r}{2}\qty(1 + \frac{\sin\qty(\pi\epsilon)}{\pi\epsilon}).
    \label{eq:r_renorm}
\end{equation}

We can use \cref{eq:r_renorm} to calculate the $r_c$ as a function of $\epsilon$ for the case with 2 layers of measurements. We know from numerical simulations that $r_{c,renorm} \approx 0.55$ which implies $r_{c}\qty(\epsilon) \approx \frac{1.1}{1 + \frac{\sin\qty(\pi\epsilon)}{\pi\epsilon}}$. Specifically, for $\epsilon = 0.5$, we find that $r_c \approx 0.68$, which is in agreement with what we find in \cref{fig:err_Rot}(a).

\subsection{Effects of Decoherence}
\begin{figure}[!t]
\includegraphics[width=.45\textwidth]{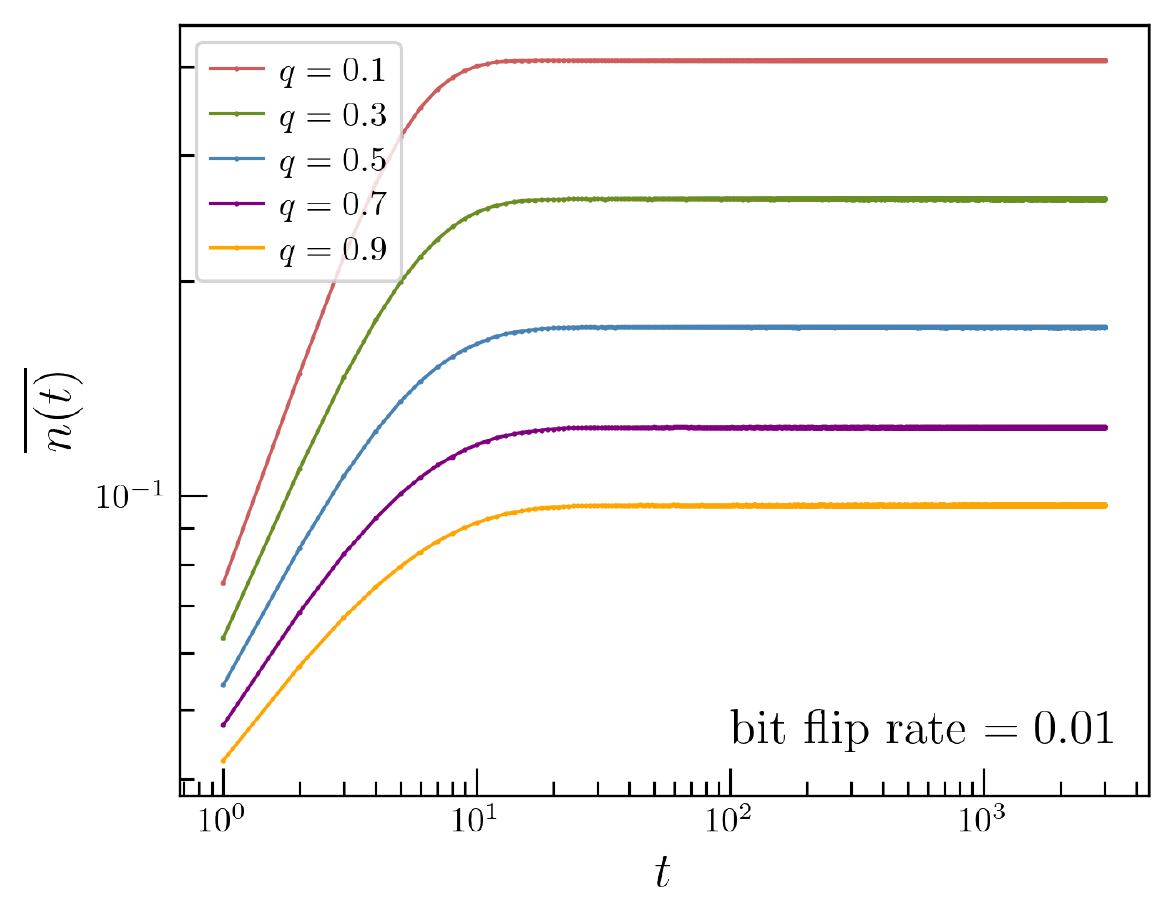}
\caption{The particle density $\overline{n(t)}$ vs $t$ for the seeding process starting with a pair of particles under bit flip errors with an error rate of 0.01. The circuit contains three sets of measurements with each set followed by a layer of random bit flip errors. There is a finite particle density in this case even at large $q$, which indicates that the absorbing state transition is unstable to bit-flip errors. Numerical simulations are performed for system size $L=300$ under OBC.}
\label{fig:3layers_seed_bit_flip} 
\end{figure}

 Since present-day quantum computing platforms are not perfectly isolated from the environment, there is a natural decay of coherence. One way of modeling this decoherence is to consider the depolarizing channel \cite{wilde_2017}, which is defined for a density matrix $\rho$ as 

\begin{equation}
    \mathcal{E}(\rho) = (1-b)\rho + b \frac{\mathbb{1}}{D}
\end{equation}

where $D = \Tr{\mathbb{1}}$ is the dimension of the Hilbert space under consideration, and $0\leq b\leq1$ represents an ``error rate". $\rho$ can refer to the density matrix of any part of the system (or the system in its entirety). In our case, we investigate the robustness of the absorbing phase transition to the effects of a depolarizing channel acting randomly on any qubit in the system. This can be implemented directly in the classical model by flipping each bit independently at a fixed rate. As shown in Fig.~\ref{fig:3layers_seed_bit_flip}, for the seeding process starting with a pair of particles, even a tiny error rate of 0.01 leads to a finite particle density in the steady state for large $q$. Therefore, the absorbing state transition is unstable to the presence of depolarizing noise.

\subsection{Effects of Off-Diagonal Dephasing}

The final error we consider is that of the loss of coherence in the off-diagonal elements of the density matrix. Since both the $ZZ$ measurements and the domain wall density $n$ share a basis, and are both diagonal in the computational basis, we do not expect this dephasing to have an impact on the observation of the order-disorder transition.

\section{Sampling Protocol}

In this section, we elaborate on our proposal for obtaining the averaged domain wall density $\overline{n(t)}$. 
Recall that the definition of $n$ is

\begin{equation}
    n \equiv \sum\limits_{j=1}^{L-1}\expval{\frac{1- Z_j Z_{j+1}}{2}}.
\end{equation}
The protocol also involves measuring $Z_j Z_{j+1}$ (and applying feedback at some rate $r$), first for odd and then even $j$ after a layer of random unitaries. In numerical simulations, we were able to directly calculate $\expval{\frac{1- Z_j Z_{j+1}}{2}}$ without any additional steps. However, in practical experiments, one would have to first prepare a state and then perform several measurements in order to estimate the expectation value of $Z_j Z_{j+1}$, which a priori poses a large overhead. Since our protocol involves measurements of this operator, we can utilize these outcomes to estimate $n$.

\begin{figure}
    \centering
    \subfigure[]{\includegraphics[width=0.485\textwidth]{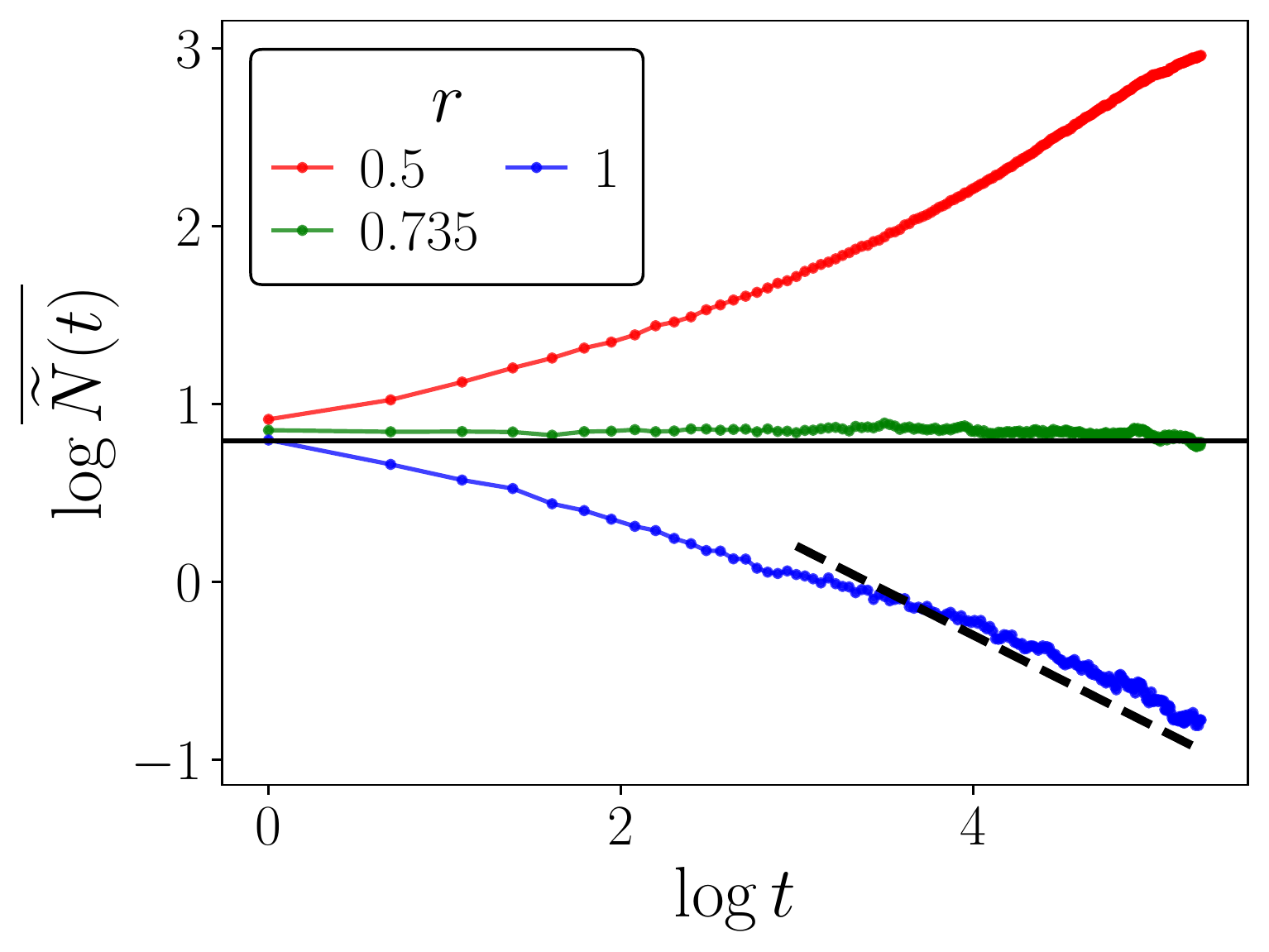}}%
    \subfigure[]{\includegraphics[width=0.5\textwidth]{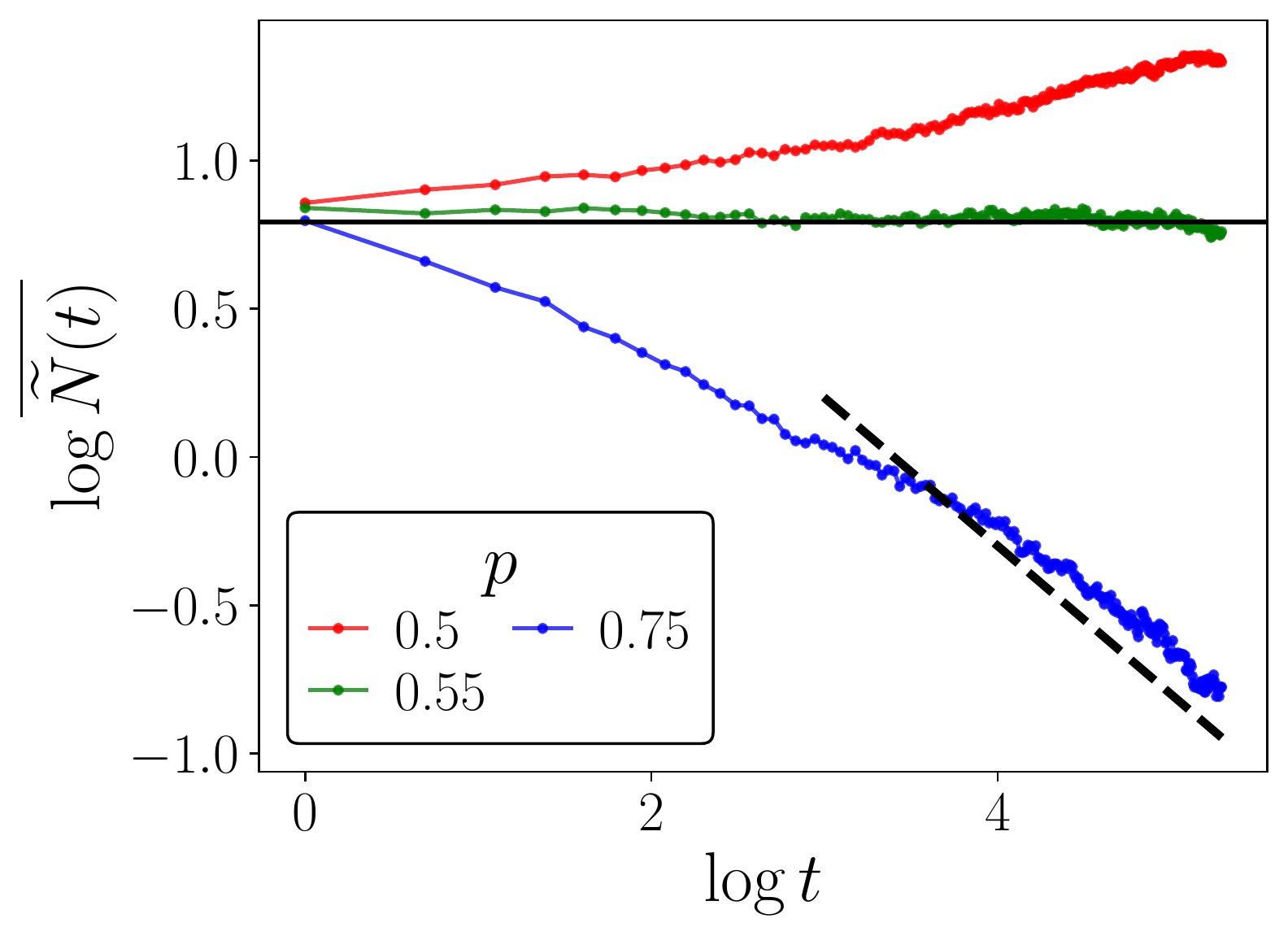}}
    \caption{The estimated number of domain walls $\overline{\widetilde{N}}$, as obtained from the sampling procedure detailed in Sec.~IV, plotted for different values of (a) the feedback rate $r$, with $p=0.75$ and (b) the measurement probability $p$, with $r=1$. The dashed line denotes diffusive decay $\widetilde{N}\sim t^{-0.5}$. In both cases, this estimator reflects the order-disorder transition at the same value of $p\times r = q_c \approx 0.55$, where $\widetilde{N(t)}\sim t^0$ at the critical point, when starting from an initial condition with 2 adjacent domain walls at the center of the chain.}
    \label{fig:sampled_N}
\end{figure}

For concreteness, we consider the set-up with 2 sets of measurements, one each after the first and second layers of random unitaries, but not the third. At each time step, the outcomes of the last set of measurements are recorded. Let the number of measurements at this time-step be $N_m$, and the measurement outcomes be $\qty{m_i}_{i=1}^{N_m}$ with $ m_i = \pm1$.  $N_m\leq (L-1)$ and on average, $N_m = p(L-1)$. The quantity $\overline{\widetilde{n}}$, defined as
\begin{equation}
    \widetilde{n} \equiv \frac{1}{N_m} \sum\limits_{i=1}^{N_m} \frac{1 - m_i}{2}
\end{equation}
averaged over trajectories and circuit realizations, provides an accurate sampling of the true domain wall density $\overline{n}$. The reason for this is that at any $p>0$, an extensive number of measurements are still being made. We can straightforwardly obtain an estimate for the number of domain walls $\overline{\widetilde{N}}$ as $\overline{\widetilde{N}}\equiv (L-1) * \overline{\widetilde{n}}$. As shown in \cref{fig:sampled_N}, the number of domain walls obtained from this sampling procedure $\overline{\widetilde{N}}$ captures the order-disorder transition, and agrees well with the true value of $\overline{n}$.

 \section{Heuristic argument of the difference between entanglement transition and steering transition}

When $r>r_c$, as we tune $p$, the quantum trajectory undergoes two transitions: the entanglement phase transition at $p_c^{EE}$ and the steering (domain wall density) phase transition at $p_c^n$. First, it is easy to show that $p_c^{EE}\leq p_c^n$ in the steady state $|\psi\rangle$. This is because when $p>p_c^n$, $|\psi\rangle$ is spanned by $|00\ldots 0\rangle$ and $ |11\ldots1\rangle$ and the entanglement entropy $S_A$ for a subsystem A is smaller than $\log 2$. 

According to previous studies on the $1+1d$ measurement induced entanglement phase transition, we expect that when $p\geq p_c^{EE}$\cite{PhysRevX.9.031009,PhysRevB.100.134306},
\begin{align}
    S_{A}=
    \begin{cases}
        \log L_A, & L_A<\xi \\
        \log \xi, & L_A>\xi
    \end{cases}
\end{align}
where $\xi$ is the correlation length for the entanglement measure and diverges at $p=p_c^{EE}$. When $p$ is slightly larger than $p_c^{EE}$, the correlation length is still quite large with $\log\xi>\log 2$. Increasing $p$ will reduce $\xi$ and eventually when $p$ is large enough, we have $\log\xi<\log 2$. This implies that there is a finite separation between $p_c^{EE}$ and $p_c^n$. 

The difference between $p_c^{EE}$ and $p_c^n$ is more obvious in higher dimensional systems with spatial dimension $d>1$. As we vary $p$, the entanglement entropy exhibits three different scaling behaviors. Here we consider a simple case where the subsystem A is d-dimensional sphere with radius $L_A$. The leading term of the entanglement entropy satisfies:
\begin{align}
    S_{A}:
    \begin{cases}
        \sim L_A^d, & p<p_c^{EE} \\
        \sim L_A^{d-1}, & p_c^{EE}<p<p_c^n \\
        <\log 2, & p>p_c^n.
    \end{cases}
\end{align}


\section{Simulation of the Volume-Law Phase}

Lastly, we address the technical and conceptual difficulties with obtaining an accurate estimate of $p^{EE}_c$. The challenges are two-fold -- (i) The rapidly burgeoning entanglement entropy as we approach the critical point or enter the volume law phase renders MPS methods (which are designed for states with low entanglement) computationally costly, and (ii) our specific family of models are particularly prone to finite-size effects owing to the absorbing nature of the steady state.

\begin{figure}[!t]
\includegraphics[width=.45\textwidth]{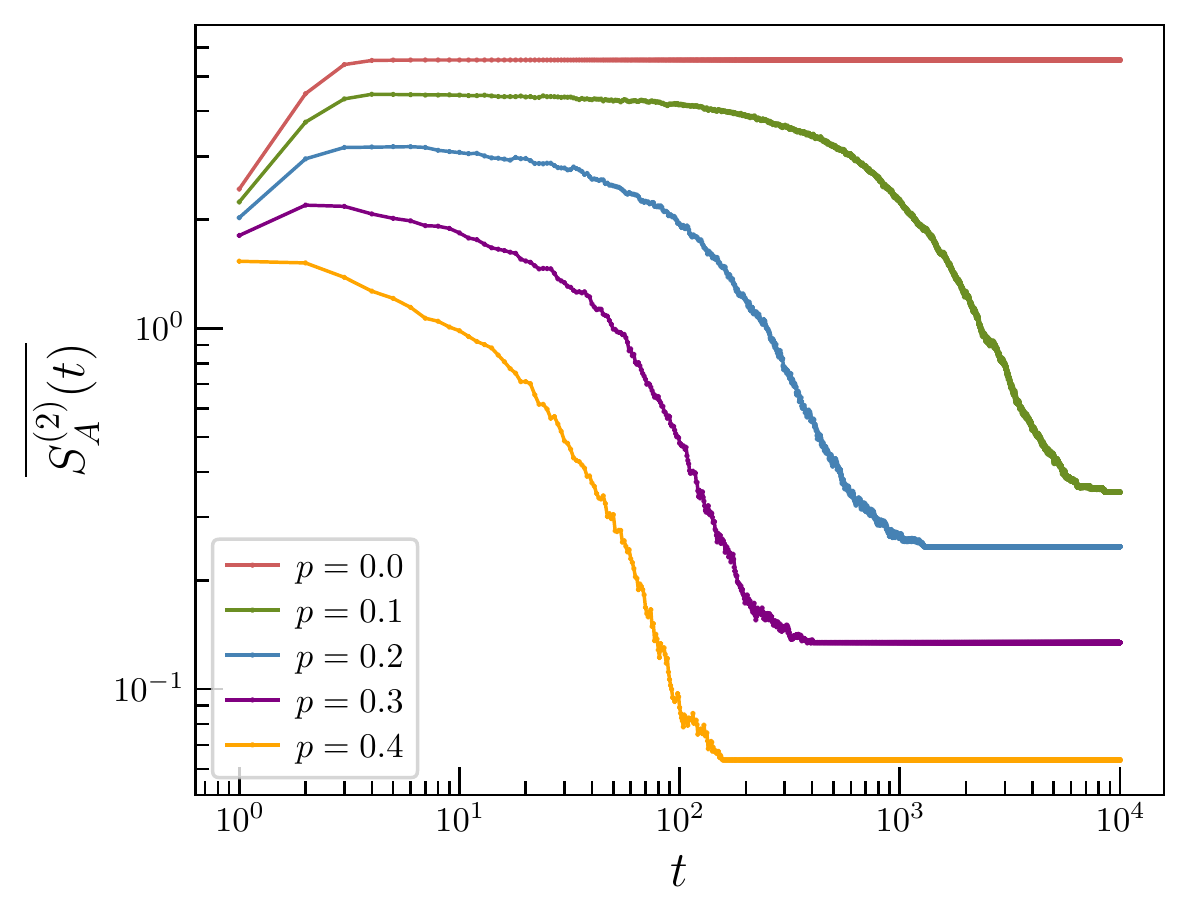}
\caption{The entanglement entropy $\overline{S_A^{(2)}(t)}$ vs $t$ plotted on a log-log scale using the exact diagonalization of the quantum trajectory under the circuit with two sets of measurements per unit time step over a variety of measurement rate $p$. All of the data are collected for system size $L=18$ and feedback rate $r=1$ under PBC.}
\label{fig:Trajectory_ED} 
\end{figure}

The volume-law phase of general (i.e. non Clifford~\cite{Aaronson_2004}) non-unitary circuits requires exponential resources and is difficult to simulate on classical computers. The exact diagonalization (ED) technique, commonly used to study the volume law phase, can only deal with small systems with $L\lesssim30$. The detrimental effects of the resulting finite-size effects are especially germane to our system. In such a small system, if $q=p\times r>0$, under BAW dynamics, any state with a finite density of domain walls quickly evolves to states with no domain walls $\{|0\cdots 0\rangle$, $|1\cdots 1\rangle\}$. The absorbing phase transition is no longer observable, since any initial configuration decays to the absorbing state. Consequently, since $p_c^{EE}< p_c^n$, the entanglement phase transition cannot be observed either in a small system with finite $r$. As shown in Fig.~\ref{fig:Trajectory_ED}, the entanglement entropy calculated using ED for system size $L=18$ with an initial product state polarized in the $+x$ direction shows a strong finite size effect and decays to a finite constant quickly for all $p>0, r=1$. This rapid decay obviates the usage of both the volume law scaling of the steady-state entanglement entropy $S^{(n)}_A(t\to\infty) \sim |A|$ and its linear growth at early times $S^{(n)}_A(t)\sim t$ as diagnostics for the volume law phase.

However, in a large system, we can observe an absorbing phase transition. In such a system,  we can use the inefficacy of MPS methods in simulating states with large entanglement to obtain an upper bound on $p^{EE}_c$ (i.e. when the infamous \lq\lq{}exponential wall" has been encountered and the bond dimension grows too large to store the state on a classical computer). For the model with two sets of measurements, we find that the upper bound for $p^{EE}_c$ is 0.45.

\bibliography{reference}